\documentclass[a4paper,11pt]{article}

\usepackage{jheppub}
\usepackage[T1]{fontenc}
\usepackage{empheq}
\usepackage{amsmath}
\usepackage{bbm}
\usepackage{amssymb}
\usepackage{dsfont}
\usepackage{graphicx}
\usepackage{hyperref}
\usepackage{stackrel}
\usepackage{comment}
\usepackage[export]{adjustbox}
\usepackage{float}
\usepackage{enumitem}
\usepackage{epsfig,amssymb,euscript}
\usepackage{amsmath,amsfonts,bbm,mathtools}
\usepackage{subcaption}
\allowdisplaybreaks[4]        
\usepackage[dvipsnames]{xcolor}
\usepackage{hyperref}
\usepackage{fancyhdr}
\usepackage[header,title,page,titletoc]{appendix}  
\usepackage[notcite,notref]{}

\DeclareMathOperator{\tr}{Tr}

\renewcommand{\(}{\left(}
\renewcommand{\)}{\right)}
\renewcommand{\[}{\left[}
\renewcommand{\]}{\right]}
\newcommand{\eg}{{\it e.g.,}\ }
\newcommand{\ie}{{\it i.e.,}\ }

\newcommand{\mt}[1]{\textrm{\tiny #1}}
\newcommand{\reef}[1]{(\ref{#1})}

\def\be{\begin{equation}}
\def\ee{\end{equation}}
\def\bea{\begin{eqnarray}}
\def\eea{\end{eqnarray}}
\newcommand{\beq}{\begin{equation}}
\newcommand{\eeq}{\end{equation}}
\newcommand{\beqa}{\begin{eqnarray}}
\newcommand{\eeqa}{\end{eqnarray}}

\def\ba{\begin{eqnarray}}
\def\ea{\end{eqnarray}}

\def\b{\beta}

\def\m{\mu}

\def\O{\Omega}
\def\r{\rho}

\def\del{\partial}
   
\begin{document}

\title{R\'enyi second laws for black holes}

\author[a,b]{Alice Bernamonti,}

\author[b]{Federico Galli,}

\author[c]{Robert C. Myers}

\author[d]{and Ignacio A. Reyes}

\affiliation[a]{Dipartimento di Fisica e Astronomia, Universit\'a di Firenze\\
Via G. Sansone 1, I-50019 Sesto Fiorentino, Italy}
\affiliation[b]{INFN, Sezione di Firenze\\
Via G. Sansone 1, I-50019 Sesto Fiorentino, Italy}
\affiliation[c]{Perimeter Institute for Theoretical Physics\\
31 Caroline Street North, Waterloo, Ontario N2L 2Y5, Canada}
\affiliation[d]{Institute for Theoretical Physics, University of Amsterdam\\
Amsterdam, 1098 XH, The Netherlands}

\emailAdd{alice.bernamonti@unifi.it}
\emailAdd{federico.galli@fi.infn.it}
\emailAdd{rmyers@perimeterinstitute.ca}
\emailAdd{ireyesraffo@gmail.com}

\abstract{Hawking's black hole area theorem provides a geometric realization of the second law of thermodynamics and constrains gravitational processes. In this work we explore a one-parameter extension of this constraint formulated in terms of the monotonicity properties of R\'enyi entropies. We focus on black hole mergers in asymptotically AdS space and determine new restrictions which these R\'enyi second laws impose on the final state. We evaluate the entropic inequalities starting from the thermodynamic ensembles description of black hole geometries, and find that  for many situations they set more stringent bounds than those imposed by the area increase theorem.}

\maketitle


\section{Introduction}

In recent years, black hole mergers have been extensively studied using numerical relativity simulations, \eg see \cite{Bantilan:2014sra,Sperhake:2009jz,Zilhao:2012gp,Zilhao:2013nda,Sperhake:2008ga,Bozzola:2022uqu,Shibata:2008rq,Healy:2015mla,Sperhake:2011ik,Sperhake:2010uv,Gold:2012tk,Sperhake:2015siy,Sperhake:2019oaw,Andrade:2019edf,Andrade:2020dgc} and \cite{Sperhake:2014nra,Choptuik:2015mma} for reviews. 
Given the complexity of Einstein's equations, an important question to ask is what can we learn about black hole evolution \textit{without} the need of solving these equations explicitly. The most important result in this regard is Hawking's area theorem \cite{Hawking:1971tu}. It states that the area $A$ of a black hole horizon can never decrease and it constrains gravitational processes such as black hole mergers. Through the insight of Bekenstein and Hawking \cite{Bekenstein:1973ur,Hawking:1974rv,Hawking:1975vcx}, this elegant geometric constraint also has an interpretation in terms of the Shannon (or von Neumann) entropy describing the underlying microphysical states of the corresponding black holes 
\begin{align}\label{BHentropy}
S=\frac{A}{4G_N}\,.
\end{align}
From this perspective, the area theorem becomes an instance of the second law of thermodynamics.

In this paper we explore the possibility that Hawking's area theorem is not an isolated monotone of the evolution, but rather one member within a larger family of laws that constrain the future state given an initial condition. Our starting point is a fundamental family of constraints that remained mostly unexplored in this context known as \textit{second laws of quantum thermodynamics} \cite{Brandao:2015qyc}. As we review in the next section, quantum systems out of equilibrium must satisfy not only the familiar second law -- that entropy increases, or free energy decreases -- but rather an infinite set of generalised second laws in their way to thermalisation.
 
A broad class of these laws, formulated in terms of the monotonicity properties of quantum R\'enyi divergences, was first explored in \cite{Bernamonti:2018vmw}  within the context of the AdS/CFT correspondence.  There, it was shown that for certain excited CFT states with a dual description in terms of a minimally coupled gravity-scalar system, R\'enyi divergences can be computed via a Euclidean path integral (see also \cite{May:2018tir}). Within this framework, it was found that there indeed exist transitions which are allowed by the traditional second law, but forbidden by the additional constraints. This showed in a toy model example, that it is possible to use the second laws of quantum thermodynamics to place more stringent restrictions on gravitational dynamics. 

Here, we explore the implications of these additional laws for classical gravity, focusing on black hole mergers in AdS and working out which new constraints they impose on the final state and on the amount of energy emitted via gravitational waves. We find that indeed for many situations they set more stringent bounds than those imposed by the area theorem. We emphasize that we do not employ in any way the holographic duality, nor holographic techniques. As we explain below, the main reason for considering black holes in AdS, rather than in flat space, comes from the fact that they can be thermodynamically stable.  
This allows us to consider a particularly neat class of additional second laws that are formulated in terms of R\'enyi entropies of the gravitational system. We refer to them as \textit{R\'enyi second laws} (or \textit{R\'enyi laws}) and evaluate them directly from the thermodynamical quantities associated to a black hole geometry in classical gravity. 
\newline

The remainder of the paper is organised as follows. Section \ref{prelim} introduces some general aspects of R\'enyi entropies and their associated second laws. In section~\ref{sec3+1}, we analyse the implications of the R\'enyi second laws in the familiar and simplest nontrivial case provided by the Schwarzschild AdS$_4$ solution in the canonical ensemble. We consider a black hole merger and work out the most stringent R\'enyi bound on the mass of the final black hole and on the amount of emitted gravitational radiation. In sections~\ref{sec:extremal} and \ref{sec:coldhot}, we extend the analysis respectively to extremal mergers in AdS$_4$ and to cold to hot mergers in higher dimensional AdS.
In section~\ref{sec:repAth}, we introduce and discuss an additional family of monotones that are related to, but distinct from R\'enyi entropies and which descend directly from a notion of entropy in the replica manifold. We then reconsider the example of section~\ref{sec3+1} and explore how these bounds differ from the R\'enyi laws. We conclude with a discussion in section~\ref{sec:discuss}. 


\section{Preliminaries} \label{prelim}

This section provides a brief overview of salient features of the new quantum second laws associated with R\'enyi entropies. We first review how R\'enyi entropies can be used to pose constraints on the dynamics of a system and discuss some general aspects in section~\ref{sec:RenyiLaws}. In particular, we review their dependence on the statistical ensemble in section~\ref{sec:inequiv} and examine those cases in which the R\'enyi constraints are redundant to the second law in section \ref{sec:noNew}. In view of the gravitational analysis in the rest of the paper, in section~\ref{sec:BHADS} we make some observations on the R\'enyi entropy of AdS black holes and on its geometric interpretation. 

\subsection{R\'enyi second laws of quantum thermodynamics}\label{sec:RenyiLaws}

One of the fundamental concepts in classical and quantum information theory is that of Shannon or von Neumann entropy
\begin{align}\label{S}
S(\rho) \equiv-\tr \left( \rho \log \rho \right)
\end{align}
associated to a normalized density matrix $\rho$. 
A closely related quantity is the relative entropy or Kullback-Leubler divergence \cite{ Kullback:1951zyt} 
\begin{align}\label{D}
D(\rho|\sigma) \equiv \tr \left( \rho\log \rho-\rho\log \sigma \right)\,,
\end{align}
which is a useful measure of the distinguishability of two probability distributions. Here $\sigma$ is usually referred to as the reference state, and $\rho$ the target state.  
Both von Neumann entropy and relative entropy satisfy a number of properties that make them fundamental in information theory (\eg see \cite{Nielsen:2012yss} for an account). The relation between the two is seen by choosing the reference state to be the uniform distribution $\mathbbm 1/d$ (where $d$ is the dimension of the Hilbert space), which yields
\begin{align}\label{DS}
D\left( \rho|\mathbbm 1/d \right)=\log d-S\left( \rho \right)\,.
\end{align}
The usual second law of thermodynamics (\ie the entropy of an isolated system never decreases) can then be recast as a monotonicity statement about relative entropy.   
Suppose that the time evolution of a system is given by a completely positive trace-preserving (CPTP) map.\footnote{
Consider a linear map $\mathcal E: L (\mathbf H) \to L (\mathbf H')$, where $\mathbf H$ and $\mathbf H'$ are the Hilbert spaces that describe the input and output systems respectively, and $L (\mathbf H)$ the set of linear operators on $\mathbf H$.  

a) The map $\mathcal E$ is called positive if $\mathcal E(X) \ge 0$ for any $X \in L (\mathbf H)$  such that $X \ge 0$. Moreover, $\mathcal E$ is completely positive (CP) if $\mathcal E \otimes \mathcal I_n$  is positive for any $n \in \mathbb N$, where $\mathcal I_n$ is the identity operator on $L(\mathbb C_n)$. 

b) The map $\mathcal E$ is called trace-preserving (TP) if $\tr[ \mathcal E(X)] = \tr[X]$ for any  $X \in L (\mathbf H)$. 
 } Such a map rules out negative probabilities by preserving the positivity of density matrices  -- even if an environment is coupled to the system -- and by conserving probability. The simplest example of a CPTP process is a unitary time evolution. It has been proven that for any CPTP evolution, the relative entropy is monotonically decreasing \cite{Lindblad:1975kmh,cmp/1103900757,Petz:2002eql}:
\begin{align}\label{mono}
D\left( \rho(t)|\sigma(t) \right)\geq D\left( \rho(t')|\sigma(t') \right)\, ,\ \ \  t<t'\, .
\end{align}
If the time evolution is such that the uniform distribution is stationary, taking this as our reference state $\sigma$, eq.~\eqref{DS} leads to 
\begin{align} 
S(\rho(t))\leq S(\rho(t'))\, ,\ \ \ t<t'\,.
\end{align}

In 1961, Alfred R\'enyi generalised these notions by introducing a \textit{family} of new entropies and divergences that share most of the fundamental properties as above \cite{renyi1961measures}. The R\'enyi entropies, which generalise eq.~\eqref{S}, are defined by
\begin{align}\label{eq:RenyiEntropy}
S_n (\r) \equiv \frac{1}{1-n} \log \tr \rho^n
\end{align}
for a normalised density matrix. Similarly, a generalisation of eq.~\eqref{D} is known as the R\'enyi divergence and is given by 
\begin{align} 
D_n\left( \rho|\sigma \right) \equiv \frac{1}{n-1} \log \tr \left( \rho^n \sigma^{1-n} \right)
\end{align}
for both $\rho$ and $\sigma$ normalised.

Although in principle R\'enyi entropies and divergences are defined for $n\in(2,3,\hdots)$, one is often interested in finding specific analytic continuations into the complex plane of $n$. If these exist, then one can show that
\begin{align} 
\lim_{n\to 1} S_n(\r)=S(\r)\qquad {\rm and}\qquad \lim_{n\to 1} D_n(\rho|\sigma)=D(\rho|\sigma)\,,
\end{align}
yielding the von Neumann entropy \eqref{S} and relative entropy \eqref{D}, respectively. There are two other interesting limits which will play a role in our discussion below: the Hartley entropy $S_0$ and the min-entropy $S_\infty$
\beq\label{Hart}
\lim_{n\to 0} S_n(\r) \equiv S_0(\r) = \log {\rm rank} (\r)\, ,\qquad \lim_{n\to \infty} S_n(\r) \equiv S_\infty (\r)= -  \log {\rm p_{max}} \, ,
\eeq
where ${\rm rank} (\r)$ is the number of nonzero elements of $\rho$, and ${\rm p_{max}}$, its maximal eigenvalue. 

R\'enyi entropies and divergences have been studied in classical and quantum information theory, quantum field theory, and holography  (\eg see  \cite{renyi1961measures,Petz:1986naj,hiai2011quantum,Wilde:2013bdg,Muller-Lennert:2013liu,Hung:2011nu,Galante:2013wta,Faulkner:2013yia,Belin:2013dva,Belin:2013uta,Belin:2014mva,Dong:2016fnf,Ugajin:2018rwd,Dong:2018lsk,Bernamonti:2018vmw,May:2018tir,Chowdhury:2019lfe,Ugajin:2020dyd,Arias:2023duc}). Viewed as an analytic function of the index $n$, R\'enyi entropies satisfy a set of inequalities (\eg see appendix A in \cite{Hung:2011nu})
\begin{align}
S_n(\r)&\geq 0\nonumber\\
\partial_n S_n(\r) &\leq 0\nonumber\\
\partial_n \left( \frac{n-1}{n} S_n(\r) \right) &\geq 0\nonumber\\
\partial_n \left[ \left( n-1 \right) S_n(\r) \right] &\geq 0\nonumber\\
\partial_n^2 \left[ \left( n-1 \right) S_n(\r) \right] &\leq 0\,. \label{ineqs}
\end{align}
Moreover, R\'enyi entropies and divergences satisfy an analogous relation to eq.~\eqref{DS}:
\begin{align}\label{DnSn}
D_n\left( \rho|\mathbbm 1/d \right)=\log d-S_n\left( \rho \right)\,.
\end{align}

It turns out that quantum systems satisfy additional constraints, which resemble the second law of thermodynamics, stemming from the properties of these divergences \cite{Brandao:2015qyc}. The main observation being that R\'enyi divergences are also monotonic 
\begin{align}\label{monon}
D_n\left( \rho(t)|\sigma(t) \right)\geq D_n\left( \rho(t')|\sigma(t') \right)\, ,\ \ \ t<t'\, ,
\end{align}
and thus provide a family of constraints along the evolution. In \cite{Brandao:2015qyc}, the authors focused on out-of-equilibrium states as they move towards thermalization. In this case, the final thermal state $\rho_\b$ is an equilibrium state, thus a fixed point of the dynamics. Using this as reference state, the monotonicity constraints then reduce to
\be \label{mononv2}
D_n\left( \rho(t)| \r_\b \right)\geq D_n\left( \rho(t')| \r_\b \right)\, ,\ \ \ t<t' \,. 
\ee
An interesting extension of the above was made in \cite{Bernamonti:2018vmw} and comes from realizing that, in fact, this equation holds for any equilibrium reference state $\r_{\rm R}$ of the system, \ie for any state which remains invariant under the evolution of interest. For example, for closed system dynamics where $\rho$ represents the full degrees of freedom of the system, all thermal states will be preserved and so we may apply eq.~\eqref{mononv2} with $\r_\b$ replaced by $\r_{\rm R}$, a general thermal state with an arbitrary temperature $1/\b_{\rm R}$. Intuitively, we can think that as the system moves in the space of states towards equilibrium, not only the distance from the specific final state $\rho_\b$ is decreasing, but also the one from the entire class of equilibrium states $\rho_{\rm R}$ available to the system. 

In this paper we would like instead to exploit an additional observation made in \cite{Bernamonti:2018vmw}. If the Hamiltonian keeps the maximally mixed state $\sigma=\mathbbm 1/d$ fixed (\ie the limit of $\r_{\rm R}$ where $\b_{\rm R}\to0$), then one can also use this as the reference state. Then, substituting eq.~\eqref{DnSn} into  eq.~\eqref{monon} implies a second law for each R\'enyi entropy
\begin{align}\label{SnSn}
S_n(\rho(t))\leq S_n(\rho(t'))\, ,\ \ \ t<t' \,. 
\end{align}
In general R\'enyi entropies do not factorise in the form $S_n = g(n) S$. Therefore the conditions \eqref{SnSn} impose constraints of different strengths for different $n$ and provide an infinite family generalisation of the second law of thermodynamics, with the usual second law corresponding to the special case $n=1$. In particular, the value $n=1$ has nothing special to it, {\it i.e.}, generically it will not be the strongest bound. The  value of $n$ that constrains the system the most will be located either at a local minimum, or at one of the boundaries, namely $n=0$, $n=\infty$, or some other value which acts as a boundary.
We will refer to the constraints \eqref{SnSn} as \textit{R\'enyi second laws}.\\

In this work, we explore the consequences of the R\'enyi second laws \eqref{SnSn} in the context of black hole physics. As described in the introduction, Hawking's area theorem (corresponding to the usual second law of thermodynamics) puts constraints on the evolution of gravitational systems involving black holes  \cite{Hawking:1971tu}. We explore how the other constraints for $n\neq 1$ put bounds on the final state of black hole mergers and on the amount of emitted gravitational radiation. 
 
For thermal systems, the uniform distribution corresponds to an infinite temperature state. As is well known, the thermodynamic and dynamical stability of these solutions depends on the asymptotic boundary conditions, ensemble, charges, etcetera. As we will argue below, black holes in AdS space possess the correct features needed for the above analysis to hold. 
 

\subsection{On the ensemble inequivalence of R\'enyi entropies} \label{sec:inequiv}

As first discussed by Boltzmann and Gibbs, in statistical mechanics and thermodynamics an important concept is that of \textit{ensemble equivalence}. This refers to the fact that, although the microstate description in different ensembles takes a very different form (either as expectation values or sharp constraints), in the strict thermodynamic limit  they all give rise to the same entropy function. This implies that the physical constraints coming from the second law of thermodynamics (\ie the monotonicity of relative entropy)  are independent of the ensemble. However, this does not hold for their R\'enyi counterparts: R\'enyi entropies and divergences are ensemble \textit{dependent}, since they capture the fluctuations or response of the system under perturbations. This means that the constraints stemming out of them will also depend on the ensemble. Implications of this observation were also discussed for chaotic systems in \cite{Lu:2017tbo} and in the holographic context in \cite{Dong:2018lsk}.\footnote{We thank Pasquale Calabrese and Erik Tonni for useful discussions on this point (see also \cite{Alba:2017kdq,Mestyan:2018fwt}).}

Let us illustrate this situation comparing the canonical and grand canonical ensembles. We  denote the partition function with, \eg $Z=\tr e^{-\beta H}$ and $Z_n=\tr e^{-n\beta H}$ for the canonical ensemble. Consider expanding the R\'enyi entropy around $n=1$: 
\bea
S_n &=& - \frac{\partial_n Z_n|_{n=1}}{Z} + \log Z  + \frac{1-n}{2 Z^{2}} \left[ Z \partial^2_n Z_n|_{n=1} - \( \partial_n Z_n|_{n=1}  \)^2 \right] + \hdots \nonumber \\
 &=&   S  +\frac{1- n}{2 Z^2 } \left[ Z \partial^2_n Z_n|_{n=1} - \( \partial_n Z_n|_{n=1}  \)^2 \right] + \hdots \,, \label{eq:dnSn}
\eea
where we used  
\be
\tr \rho^n \equiv \frac{Z_n}{(Z)^n} \,.
\ee

The point we wish to emphasize here is that while $S$ is independent of the ensemble, the higher order terms are \textit{not}, even in the thermodynamic limit. As we show below, the reason is that while $S$ depends on mean value quantities such as the energy $\langle H\rangle$ or charge $\langle Q\rangle$, the R\'enyi entropies contain information about \textit{fluctuations}, which are ensemble dependent. Since, as we reviewed above, R\'enyi entropies are special cases of R\'enyi divergences, we  expect that in general, the divergences also differ for different ensembles.

Consider a system with Hamiltonian $H$ and charge $Q$. The associated potentials are the inverse temperature $\beta$ and the chemical potential $\mu$. In the canonical ensemble, we fix $\beta$.   
The partition function is $Z^{\rm c}(\beta)=\tr  e^{-\beta H} $, and therefore the von Neumann entropy in this ensemble reads
\begin{align}\label{Sc1}
S^{\rm c}=\beta \langle H\rangle + \log Z^{\rm c}(\beta)\,.
\end{align}
For the grand canonical instead, the partition function is $Z^{\rm gc} (\beta,\mu)=\tr e^{-\beta H+\mu Q}  $ and its entropy is
\begin{align}\label{Sg1}
S^{\rm gc}=\beta \langle H\rangle - \mu \langle Q\rangle + \log Z^{\rm gc} (\beta,\mu) \, .
\end{align}
Now, the ``equivalence of ensembles'' refers to the equality between eqs.~\eqref{Sc1} and \eqref{Sg1} when we take the thermodynamic limit, which assumes homogeneity, a macroscopic limit and no phase transitions. Under these assumptions, one can show that 
\cite{1987stme.book.....H} 
\begin{align}\label{equiv}
\log Z^{\rm gc} (\beta,\mu)=\log Z^{\rm c}(\beta)+\mu \langle Q\rangle \,,
\end{align}
and substituting into eq.~\eqref{Sg1}, we immediately recover eq.~\eqref{Sc1}. Therefore the Shannon entropy is equal in the two ensembles. In what follows we will always assume to work in the thermodynamic limit and thus $S^{\rm c} = S^{\rm gc} \equiv S$. 

Moving now to the first correction in eq.~\eqref{eq:dnSn}, we can again compute it in both ensembles and compare the results. In the canonical ensemble, we find
\be
S_n^{\rm c} = S + \frac{1-n}{2} \b^2 \left( \langle H^2 \rangle - \langle H \rangle^2 \right)\,,
\ee
whereas in the grand canonical ensemble
\be
S_n^{\rm gc} = S + \frac{1-n}{2} \left\{ \b^2 \left( \langle H^2 \rangle - \langle H \rangle^2 \right) + \mu^2  \left( \langle Q^2 \rangle - \langle Q \rangle^2 \right) - 2 \b  \mu \(\langle H Q \rangle - \langle H \rangle \langle Q \rangle\) \right\}\,.
\ee
Thus while the entanglement entropies (with $n=1$) of the two ensembles are equal, the R\'enyi entropies are not since they differ in their higher order terms. Of course, we see that the difference vanishes if the chemical potential $\mu$ is zero. 

This inequivalence of ensembles for R\'enyi entropies will become important later on when we study black hole collisions, where we will see that processes with initial and final states in one or another ensemble will give rise to different bounds - 
see further discussion in sections \ref{ReplicaEnsembles} and  \ref{sec:discuss}.

\subsection{When are there no new constraints?}\label{sec:noNew}

In general R\'enyi entropies do not factorise in the form $S_n = g(n)\,S$. There are however surely two instances in which the R\'enyi laws are redundant with respect to the ordinary second law: in the microcanonical ensemble and whenever the entropy is a homogeneous function of the charges of the system. We discuss the two cases below. 


\subsubsection{Microcanonical ensemble}

In the microcanonical ensemble, the state of the entire system is characterized by a fixed total energy within a narrow band $[E, E+ \Delta E]$, with $\Delta E \ll E$. The normalized density matrix can be written as
\be
\rho = \frac{1}{\O} \, \left\{ \Theta(H-E) - \Theta (H-E -\Delta E) \right\}\,,
\ee
where $\O$ denotes the total number of microstates and $\Theta$,  the Heaviside step function. For the R\'enyi entropies \eqref{eq:RenyiEntropy} of the entire system, we thus have in the thin energy shell limit
\be
\tr \rho^n \approx \O^{1-n}
\ee
and 
\be \label{snmc}
S_n^{\rm mc} = \log \O \equiv S\,.
\ee
All R\'enyi entropies are equivalent to the ordinary entropy (see also eq.~(22) in \cite{Dong:2018lsk} with $f=1$), and do not provide additional laws. 


\subsubsection{Homogeneous entropy functions}\label{sec:homentropy}

Consider now a system with various conserved charges, say $M$ and $Q$ and their associated potentials $\beta$ and $\mu$, although the discussion is identical for any number of charges. 
Suppose we already solved for the entropy $S(M,Q)$ as a function of the charges. If the entropy is a homogeneous function of the charges:
\begin{align}\label{sca}
S(\lambda M,\lambda Q)=\lambda^\nu S(M,Q)\,,
\end{align} 
then we can show that the grand canonical R\'enyi entropies factorise in the form
\begin{align}\label{Sn2}
S^{\rm gc}_n&= \frac{n }{1-n} \left( 1- n^{\frac{1}{\nu-1}}\right) \left( \nu-1 \right)S
\end{align}
and therefore the R\'enyi second laws are redundant.

To see this let us consider the grand canonical ensemble  
\be
Z^{\rm gc}(\beta,\mu) = \tr e^{-\beta H+\mu Q} 
\ee
and the corresponding entropy  \eqref{Sg1}
\be \label{SMQ} 
S  
= \beta  \langle H \rangle - \mu  \langle Q \rangle + \log Z^{\rm gc} (\beta,\mu) \, .
\ee 
Knowing  $S $ as a function of  $ \langle H \rangle  = \tilde M$ and  $\langle Q \rangle= \tilde Q$, the corresponding potentials are given by the usual thermodynamic equations 
\begin{align}\label{sad}
\beta=\frac{\partial}{\partial M} S(M,Q) \,,\qquad \mu=-\frac{\partial}{\partial Q} S(M,Q)\,. 
\end{align}
In this case, one inverts eq.~\eqref{sad} to find $\tilde M(\beta,\mu),\tilde Q(\beta,\mu)$ and substitute them back into eq.~\eqref{SMQ} to find
\be\label{I}
\beta F(\beta,\mu) =- \log Z^{\rm gc} (\beta,\mu) =\left( \beta \tilde M-\mu \tilde Q-S(\tilde M,\tilde Q) \right)\Big|_{\tilde M(\beta,\mu),\tilde Q(\beta,\mu)} \, .
\ee
Now, one can directly compute the Renyi entropies as
\begin{align} 
S_n^{\rm gc}&=\frac{1}{1-n}\log \frac{\tr\ e^{n(-\beta H+\mu Q)}}{\left[ \tr\ e^{-\beta H+\mu Q} \right]^n}\nonumber  \\
&=\frac{1}{1-n}\log \frac{Z^{\rm gc}(n\beta,n\mu)}{\left[ Z^{\rm gc}(\beta,\mu) \right]^n}\nonumber \\
&=\frac{ n \beta }{1-n}\left[ F(\beta,\mu)-F(n\beta,n\mu) \right]\label{renyi}\, .
\end{align}
If however, the entropy function satisfies a scaling relation as in eq.~\eqref{sca}, then the calculation simplifies. 
Indeed, to find $F(n\beta,n\mu)$ we need to solve eq.~\eqref{sad} but now for $n\beta$ and $n\mu$. But this is easily achieved: whatever the solution to the system \eqref{sad} is, then the charges $\tilde M'=\lambda \tilde M,\tilde Q'=\lambda \tilde Q$ will be a solution to the system 
\begin{align} 
n\beta&=\frac{\partial}{\partial (\lambda M)}S(\lambda M,\lambda Q)=\lambda^{\nu-1} \beta\\
n\mu&=\frac{\partial}{\partial (\lambda Q)}S(\lambda M,\lambda Q)=\lambda^{\nu-1}\mu
\end{align}
provided we choose
\begin{align} 
\lambda=n^{\frac{1}{\nu-1}}\,. 
\end{align}
Thus, we have that
\begin{align}
n \beta F(n\beta,n\mu)=n\beta \lambda \tilde M-n\mu \lambda \tilde Q-S(\lambda \tilde M,\lambda \tilde Q)=n^{\frac{\nu}{\nu-1}} \beta F(\beta, \mu)\, .
\end{align}
Finally, substituting back into the R\'enyi entropy \eqref{renyi}, we find
\begin{align} 
S_n^{\rm gc} 
=\frac{n \beta }{1-n}\left( 1-n^{\frac{1}{\nu-1}} \right)  F(\beta,\mu)
\end{align}
where all $n-$dependence has factored out.  
One can now take the limit $n\rightarrow 1$, which gives
\begin{align} 
S&=\frac{\beta}{\nu-1}F(\beta,\mu)\,,
\end{align}
and as claimed 
\begin{align}\label{eq:Sn}
S_n^{\rm gc}&=\frac{n }{1-n} \left( 1- n^{\frac{1}{\nu-1}}\right) \left( \nu-1 \right) S\,. 
\end{align}
More in general the argument holds whenever the entropy is a homogeneous function of the variables conjugate to those appearing in $Z$. 
 

\subsection{Black holes in Anti de Sitter} \label{sec:BHADS}

Black holes in AdS space have several remarkable features that distinguish them from black holes in asymptotical flat space. One of these features is that AdS black holes can be thermodynamically \textit{stable}: their heat capacity is positive, and they are in thermal equilibrium with their own radiation. Their temperature is an increasing function of their size. On the contrary, black holes in flat space are generically not stable: their temperature increases as they get smaller, and they eventually evaporate. 

As we argued above, for an evolution that has the maximally mixed state as a fixed point, the monotonicity of R\'enyi divergence under CPTP maps implies the monotonicity of R\'enyi entropies \eqref{SnSn}. In the case of AdS black holes, any black hole with temperature (or horizon radius) beyond the Hawking-Page transition is stable in the canonical ensemble \cite{Witten:1998zw} and therefore a fixed point. In particular, the infinite temperature black hole is stable, and so the argument applies. On the other hand, it can be shown that for negative heat capacity, such as is the case for black holes in flat space, the fundamental R\'enyi inequalities \eqref{ineqs} are generically violated outside the trivial microcanonical ensemble, and the argument fails.

In the following sections we evaluate the R\'enyi entropies for a number of AdS black holes, and verify explicitly these satisfy the properties \eqref{ineqs}. For comparison, in appendix \ref{app:flat}, we consider the same quantities for Schwarzschild black holes in flat space and verify these instead violate the inequalities  \eqref{ineqs}. 


\subsubsection{Homogeneous black hole entropy}

Before moving on to a detailed analysis, let us first make a preliminary observation in view of the discussion of section \ref{sec:homentropy} applied to AdS black holes. Interestingly, it is not necessary to solve the equation for the horizons and compute explicitly the entropy to find eq.~\eqref{sca}. 
Rotating BTZ is one example \cite{Banados:1992wn}. The horizon equation is 
\begin{align}\label{BTZf}
f(r)=- 8 M G_N+\frac{r^2}{\ell^2}+\frac{16 G_N^2 J^2}{r^2}=0
\end{align}
in terms of the ADM mass $M$, angular momentum $J$ of the solution and AdS radius $\ell$. 
We see that $M\rightarrow \lambda M,J\rightarrow \lambda J$ and $r\rightarrow \sqrt \lambda\ r$ leave the equation invariant. Given that $S(M, J) \sim r_+ $, this implies that
\begin{align} 
S(\lambda M,\lambda J)=\lambda^{1/2}\, S(M,J)\,.
\end{align}
So $\nu=1/2$ here and we find the R\'enyi entropy from eq.~\eqref{eq:Sn},
\begin{align}\label{eq:RenyigcBTZ}
S_n^{\rm gc}= \frac{n+1}{2n}\, S\,. 
\end{align}
This function satisfies all required properties \eqref{ineqs}.  

Notice this is not true for rotating black holes in higher dimensions. In Kerr-AdS$_4$ for instance the entropy is given by \cite{Hawking:1998kw}
\be
S = \pi  \, \frac{r_+^2 + a^2}{G_N \Xi}\,,
\ee
with $J = a M$, $\Xi = 1- \frac{a^2}{\ell^2}$ and $r_+$ denoting the outer event horizon, \ie the larger root of 
\be
\Delta(r) = (r^2 + a^2) \(1+ \frac{r^2}{\ell^2}\) - 2 m r = 0\,,
\ee
with $m = M G_N \Xi^2$. Under $M \to \lambda M, J \to \lambda J$, $a$ and $\Xi$ remain invariant, but $r_+$ is modified and thus the entropy is not a homogeneous function of the charges. 


\subsubsection{Replica ensembles}\label{ReplicaEnsembles}

In the rest of our analysis, we will evaluate R\'enyi entropies for AdS black holes associated with states $\rho$ in different ensembles. We will mainly use the analytic properties of the entropy and other thermodynamical quantities directly associated with $\rho$,  as in eq.~\eqref{renyi}, rather than explicitly constructing $\rho^n$.
 It is however also possible to associate a geometric meaning to $\rho^n$, or more precisely to the normalized replica density matrix
\begin{align} 
 \rho_n \equiv \frac{\rho^n}{\tr  \rho^n }\,  
\end{align}
and the corresponding von Neumann entropy\footnote{The identity  follows directly from the definitions for $\rho$ normalised so that $\tr \rho=1$.}
\begin{align}\label{obs}
S \left( \rho_n \right) &= n^2 \partial_n \left( \frac{n-1}{n} S_n\left( \rho \right) \right) \equiv    \tilde S_n\left( \rho \right) \, , 
\end{align}
which was dubbed refined entropy  \cite{Dong:2016hjy,Dong:2018lsk} and differs from the R\'enyi entropy $S_n(\rho)$.   
This also yields a direct geometric  picture for the inequivalence of  ensembles for Renyi entropies.

To illustrate this, consider a Euclidean black hole geometry characterized by a certain inverse temperature $\beta$ and ADM energy $M$.  
Its geometric entropy associated with the horizon area can equivalently be interpreted as the von Neumann entropy in the microcanonical ensemble with energy $E=M$, or in the canonical ensemble with inverse temperature $\beta$, where $\langle H \rangle = M $. 
When considering  $\rho_n$ this choice has however a clear geometric manifestation. To illustrate this we will follow \cite{Dong:2018lsk}. They  worked in the holographic context and considered a portion of the system, but the argument adapts straightforwardly to our situation (see also \cite{Hung:2011nu,Faulkner:2013yia}).

The gravitational geometry $\rho_n$ can be described in terms of the Euclidean $\mathbb Z_n$-symmetric replica manifold $\mathcal{M}_n$.\footnote{An alternative description is in terms of the quotient manifold  $\mathcal{M}_n/ \mathbb Z_n$ . For smooth solutions with a horizon representing a $\mathbb Z_n$ fixed point in $\mathcal{M}_n$, the quotient has a bulk defect with opening $2\pi/n$ in the time direction \cite{Dong:2018lsk}. 
}
 In the situation at hand, one should look for asymptotically AdS solutions of the vacuum Einstein's equations, which are smooth in the bulk interior and satisfy the asymptotic boundary conditions on $\partial \mathcal{M}_n$ specific to the desired ensemble.
 The canonical ensemble  is obtained imposing a periodicity $n \beta$ for the Euclidean time direction on the boundary. In  the microcanonical description one instead needs to choose the time periodicity as to keep the energy fixed to $E$ for any choice of $n$.
 At high enough temperatures (above the deconfinement transition of the replicated manifold for any given $n$), the geometry has a horizon with an associated entropy given by
\begin{align} 
S(\rho_n) =\frac{A_n}{4G_N} \, ,
\end{align}
which corresponds to the  von Neumann entropy of $\rho_n$ and is equal to $\tilde S_n\left( \rho \right)$ in eq.~\eqref{obs}.\footnote{Notice that $S(\rho_n)$ interpreted as the entropy  for the replicated system retains its characteristic of being ensemble independent. That is, $\rho_n$ obtained as the replica of $\rho$ in the (say) canonical ensemble with inverse temperature $\beta$ could also be interpreted from the point of view of the replica manifold as describing a microcanonical ensemble with energy matching $\tr\( H \rho_n \)$. This however in general differs from the microcanonical ensemble one associates to the original $\rho$.}

A simple concrete example is obtained considering the BTZ black  hole  eq.~\eqref{BTZf}, with  $J=0$ and horizon radius $r_+$. This has \cite{Banados:1992wn}
\be
\beta  = \frac{ 2\pi \ell^2}{r_+}\,,  \qquad  M = \frac{r_+^2}{8 \pi G_N \ell^2}
\ee
and its geometric entropy
\be
S = \frac{r_+}{4 G_N}
\ee
can be associated either with a microcanonical or canonical statistics. 

When considering the replicas $\rho_n$,  the canonical ensemble geometry will be a Euclidean BTZ solution  with inverse temperature $n \beta$, which corresponds to a smooth horizon at  $r_+ /n$ and a rescaled mass $M/n^2$. The refined entropy  in this case is 
\be
\tilde S^{\rm c}_n(\rho) =  \frac{1}{n}S(\rho)
\ee
in agreement with eq.~\eqref{obs} and the fact that $S_n^{\rm c}(\beta)= \frac{n+1}{2n} S(\beta)$ (see appendix \ref{app:AdS3}).

Given the condition on the energy,   the microcanonical geometric description for $\rho_n$ turns out to be given by the same regular Euclidean BTZ geometry one started from, and 
\be
\tilde S^{\rm mc}_n(\rho)  =  S(\rho)\,,
\ee
 as expected from substituting eq.~\eqref{snmc} into eq.~\eqref{obs}.
 

\section{Four-dimensional Schwarzschild AdS}
\label{sec3+1}

We first illustrate the implications of the R\'enyi second laws by considering a familiar example, Schwarzschild black holes in four-dimensional AdS space and working in the canonical ensemble. This is also the easiest nontrivial case since no new constraints arise for three dimensions (\ie the BTZ black holes) in either the canonical or grand canonical  ensemble.\footnote{However, we present a nontrivial three-dimensional example in appendix \ref{app:AdS3} by introducing a nonstandard ensemble.} While it is relatively straightforward to extend the analysis below to Schwarzschild AdS solutions in higher dimensions (see appendix \ref{app:SAD}), the following discussion with four dimensions will illustrate the salient results.


\subsection{R\'enyi entropies in the canonical ensemble}\label{sec:RenyiCanonical}

\begin{figure}[t]  
\begin{center}
\includegraphics[width=0.5\textwidth]{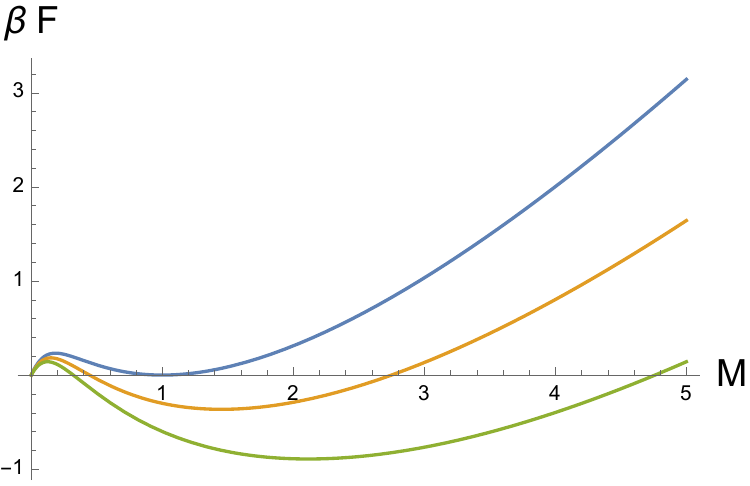}
\caption{$\beta F = \beta M - S(M)$ as a function of $M$ for $b=\frac{\beta}{2\pi \ell} =0.404, \ 0.452,\ 0.500 $, beginning with the bottom (green) curve and moving upwards.}
\label{fig:AdS4}
\end{center}
\end{figure}

In practice, determining the R\'enyi entropies analytically depends on whether one can write down the free energy $F(\beta)$ in a closed form in terms of the temperature. Indeed, for a thermal density matrix
\be \label{eq:thermalrho}
\rho = \frac{e^{-\b H}}{Z^{\rm c}(\b)}\,,
\ee
the R\'enyi entropy $S_n$ can be written in the canonical ensemble as \cite{Baez:2011upp,Hung:2011nu} 
\bea
S_n^{\rm c} &=& \frac{1}{1-n} \log \tr \rho^n \nonumber \\
&=& \frac{n \b}{1-n} \(F(\b) - F(n \b)\) \label{eq:Snsecond}
\eea
in terms of the free energy
\be
F(\b) \equiv - \frac  1 \b \log Z^{\rm c}(\b)\,.
\ee
Given that the entropy is
\be
S(\b) = \b^2 \frac{\del F}{\del \b}\,,
\ee
we can then compute $S_n^{\rm c}$ from  \cite{Baez:2011upp,Hung:2011nu} 
\bea \label{eq:deltaF}  
F(\b) - F(n \b) = \int_{n \b}^\b \frac{d\tilde \b}{\tilde \b^2} \ S(\tilde \b)\,. 
\eea
We now evaluate explicitly this expression in the case of Schwarzschild AdS$_4$
\be
ds^2 = - f(r)\, dt^2 + \frac{dr^2}{f(r)} + r^2 \,d\O_2^2 \,.
\ee
In the following, we set $G_N =1$ and then the blackening factor for this black hole geometry becomes
\begin{align}\label{f}
f(r)=1-\frac{2M}{r}+\frac{r^2}{\ell^2} 
\end{align}
in terms of the ADM mass $M$ and AdS radius $\ell$. The real root to $f(r) = 0$ gives the location of the event horizon $r= r_+$, in terms of which the black hole entropy and Hawking temperature read
\bea  
S&=& \pi r_+^2  \\
T&=& \frac{1}{4\pi } \left( \frac{1}{r_+}+\frac{3r_+}{\ell^2} \right) \,. \label{eq:TSch4}
\eea
Considering the function $\beta F = \beta M - S(M)$ depicted in figure \ref{fig:AdS4}, we can study the phase diagram. The extrema correspond to the possible black holes. Of course, the larger solution is stable at high enough temperatures (\ie corresponds to a local minimum), whereas the smaller is always unstable (\ie is a local maximum). 
We focus only on the stable black holes and to temperatures above the Hawking-Page critical temperature $T_\mt{HP}=\frac{1}{\ell \pi}$. The latter corresponds to a critical horizon radius for the large black holes of $r_\mt{HP}=\ell$. 

To express the free energy as a function of the temperature, we invert eq.~\eqref{eq:TSch4} and take the larger root corresponding to the large black holes
\begin{align}\label{rp}
r_+=\frac{\ell}{3 b}\left( 1+\sqrt{1-3 b^2} \right)\,,
\end{align}
where for convenience we introduced  the dimensionless inverse temperature
\begin{align}\label{dimb}
b \equiv \frac{\b}{2\pi \ell}\, ,
\end{align}
in terms of which the critical temperature at the Hawking-Page transition  corresponds to $b_\mt{HP}=1/2$. 

Eq.~\eqref{rp} can be substituted into the entropy to obtain  
\begin{align}\label{S(T)}
S(\beta)&=\frac{\pi \ell^2}{9 b^2} \left(1+\sqrt{1-3 b^2}  \right)^2\,.
\end{align}
Using eq.~\eqref{eq:deltaF}, we obtain the R\'enyi entropy of the Schwarzschild  AdS$_4$ black hole
\be\label{S_n}
S_n^{\rm c} (\b) = - \frac{\pi \ell^2}{3} \left\{ 1+ \frac{2}{9 b^2 n^2 (1-n)}  \left[ n^3 - 1 + \left(1- 3 b^2 \right)^{3/2} n^3 - \left(1- 3 b^2 n^2 \right)^{3/2} \right] \right\}\,. 
\ee

The  Schwarzschild AdS$_4$ black hole is thermodynamically stable above the Hawking-Page temperature $T_\mt{HP}$, the partition function is well defined, and the black hole is indeed a minimum of the action. Therefore the states correspond to a genuine Boltzmann probability distribution, which ensures that the R\'enyi entropies are well defined. Indeed one can check that the expression above satisfies the five inequalities given in eq.~\eqref{ineqs}. Additionally, it is easy to check that the limit $n \to 1$ of eq.~\eqref{S_n} gives back the correct thermal entropy \eqref{S(T)}. For $n \to  0$, $S_{n}$ gives the logarithm of the size of the support of the density matrix, as noted in eq.~\reef{Hart}. In the present case, this limit yields $S_n^{\rm c} \simeq \frac{4\pi\ell^2}{27b^2\,n^2} \to \infty$. Thus, although the black hole entropy is finite, the number of bits required by a memory to store the probability distribution is infinite.\footnote{When the probability distribution is stored in an approximate way, such as with a density matrix close in trace distance to the original one, this can be made finite.}

Notice that, for a fixed $\b$, the expression for $S_n^{\rm c} (\b)$  in eq.~\eqref{S_n} holds only up to $n = 1/(\sqrt 3 b)$, where the expression becomes complex. In fact, our analysis actually ceases to be valid even before that. This happens  at a critical $n$ where there is a phase transition in the R\'enyi entropies, because we reach the Hawking-Page critical point of the replicated manifold. Indeed, the gravity partition function $Z^{\rm c}(\beta)$ undergoes a first order phase transition at the critical temperature $T_\mt{HP}$, when the horizon radius and AdS radius are equal. Since the R\'enyi entropies involve computing $Z^{\rm c}(n\beta)/Z^{\rm c}(\beta)$, even if we start with a value of $\beta$ well below the Hawking-Page transition, $Z^{\rm c}(n\beta)$ will undergo that transition at $\beta=\beta_\mt{HP}/ n=\pi\ell/n$. This translates into the fact that our calculations in terms of black hole thermodynamical variables is reliable up to $n_\mt{HP} = b_\mt{HP}/b = 1/(2b) < 1/(\sqrt 3 b)$. To extend our calculation beyond this value, one would need to consider instead the dominant saddle for temperatures $T < T_\mt{HP}$, \ie thermal AdS. The R\'enyi entropy in this regime could then be evaluated from the AdS Euclidean action with time periodicity $\b$  (see also Outlook paragraph in sec.~\ref{sec:discuss}).


\subsection{Black hole mergers: bounds on the final mass} \label{sec:merM}

We consider a head-on collision of two black holes in AdS which merge into a unique black hole of mass $M_f$. 
The relation between the mass and temperature of large black holes follows from eq.~\eqref{f} 
\begin{align}   
M&=\frac{r_+}{2} \left( 1+ \frac{r_+^2}{\ell^2} \right) \nonumber  \\
&=\frac{\ell}{27 b^3}\left(1+\sqrt{1-3 b^2} \right) \left( 1+ 3 b^2 + \sqrt{1-3 b^2} \right) \,, \label{M}
\end{align}
where we substituted eq.~\eqref{rp} to produce the second expression. For a given $M$, the corresponding black hole temperature is the unique real root of the above equation. 

The entropy \eqref{S(T)} and R\'enyi entropies \eqref{S_n} only depend on the inverse temperature. We assume that the initial conditions for the merger are such that the density matrices of the two colliding black holes approximately factorize $\rho \approx \rho_1\otimes \rho_2$. One could imagine for instance the colliding black holes were produced by two independent gravitational collapses of two lumps of matter which were not originally correlated. Thus initially $S_n^{\rm c} \approx  S_n^{\rm c}(M_{1i}) +   S_n^{\rm c}(M_{2i}) $ and 
the R\'enyi bounds \eqref{SnSn} on the mass of the final state take the form 
\be
S_n^{\rm c}  (M_f)\geq  S_n^{\rm c}(M_{1i}) +   S_n^{\rm c}(M_{2i})\,. 
\ee

\begin{figure}[h]
\begin{center}
\includegraphics[width=0.45\textwidth]{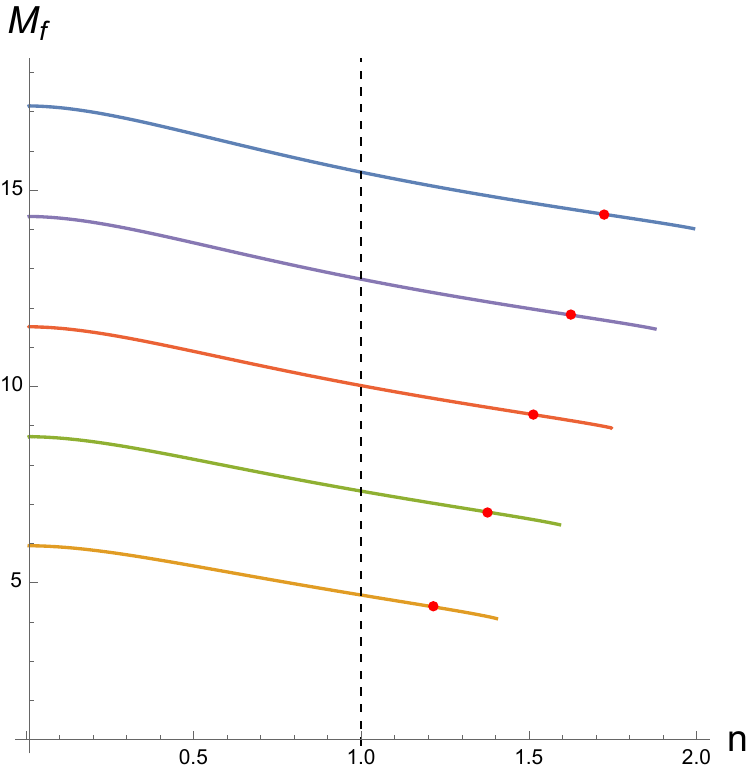}\hfill \includegraphics[width=0.45\textwidth]{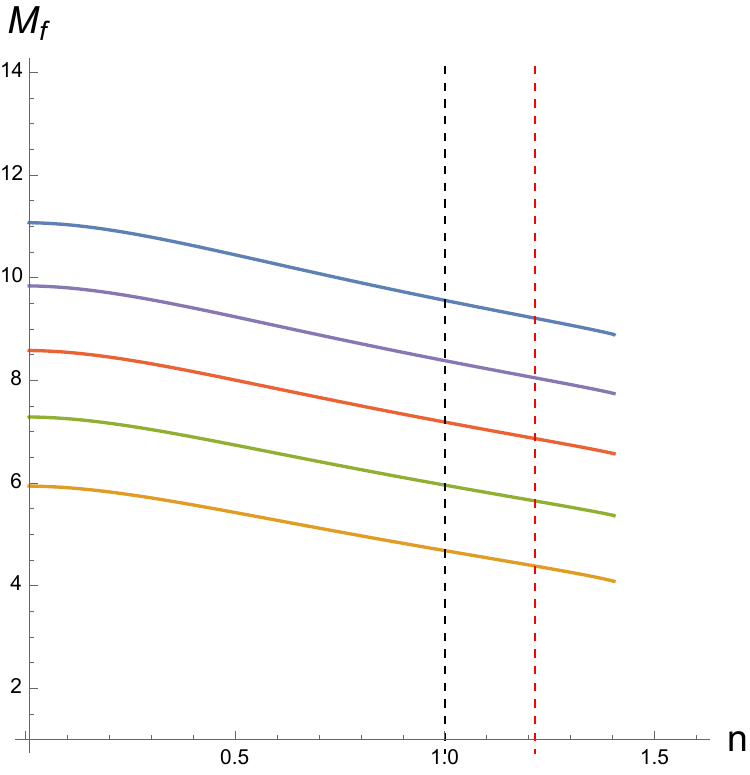} 
\end{center}
\caption{Lower bounds on the final mass $M_f$ of a head-on collision in AdS$_4$, as function of the Renyi index $n$. On the left, $M_i = 2,3,4,5,6$ from the bottom up. (We have set $\ell =1$ in both plots.) Each curve is plotted up to $n = 1/(\sqrt 3 b_i)$, where the expression \eqref{S_n} becomes complex. The red dot indicates the value $n_\mt{HP} = 1/(2b_i)$ where a phase transition in the R\'enyi entropies occurs. The expression for $S_n(M_i)$ following from eq.~\eqref{S_n}  is correct only up to that point.  
On the right, $M_{1i} = 2$ and $M_{2i} = 2,3,4,5,6$   from the bottom up. All curves stop at the same value of $n$ corresponding to the point where $S_n(M_{1i})$ becomes complex.  The dashed red line  corresponds to the value   $n_\mt{HP}$ where  $S_n(M_{1i})$  undergoes a phase transition. Beyond that value the expression for  $S_n(M_{1i})$ in  eq.~\eqref{S_n}  ceases to be valid.}\label{fig:boundM}
\end{figure}

For simplicity in the rest of the discussion, we focus on the case where the initial black holes have identical masses $M_i$. The case with different initial masses is qualitatively similar (\eg see the right panel of figure \ref{fig:boundM} for a comparison). 
We thus consider the R\'enyi inequalities \eqref{SnSn}
\be\label{another}
S_n^{\rm c}  (M_f)\geq 2 S_n^{\rm c} (M_i)
\ee
which leave one free parameter $M_f$ in terms of $n$. In the left panel of figure \ref{fig:boundM}, we plot the lower bounds on the final mass $M_f$ of a head-on collision in AdS$_4$ as function of the R\'enyi index $n$. The strongest bound comes from $n=0$ (the Hartley entropy), while Hawking's area theorem corresponds to $n=1$. The different curves correspond to varying the initial black hole masses $M_i$. 

The ordinary second law $S (M_f) \ge 2 S(M_i)$ implies 
\be
r_{f} \ge \sqrt 2 r_{i}
\ee
for the final $r_f$ and initial $r_i$ radii of the event horizons. Thus, we derive a constraint on the final mass $M_f$ from eq.~\eqref{M}, 
\be \label{eq:Mfn1}
M_f \ge M(r_+ = \sqrt 2 r_{i}) \,.
\ee

\begin{figure}[h]
\begin{center}
\includegraphics[width=0.6\textwidth]{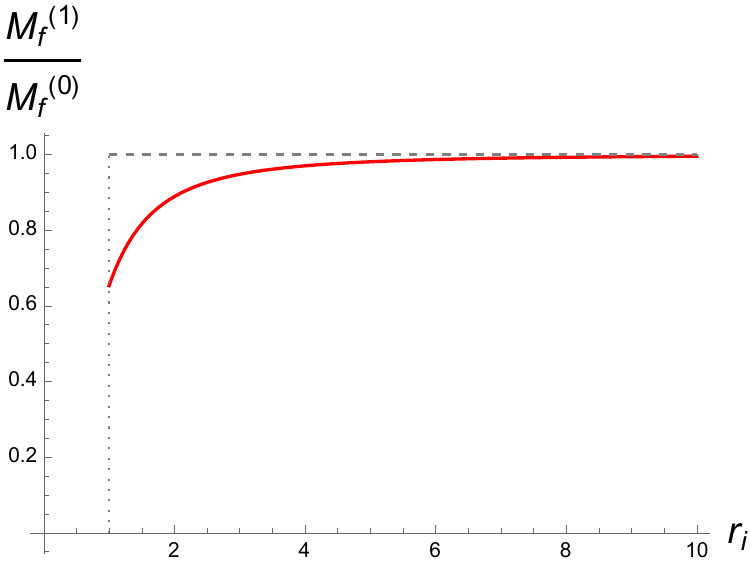} 
\caption{Ratio of the $n=1$ and $n=0$ bounds on the mass of the final state as a function of $r_{i}$. Here $\ell =1$. The dotted vertical line marks the Hawking-Page transition at $r_i = \ell$ for the colliding black holes.}\label{fig:boundMratio}
\end{center}
\end{figure}

However, the strongest bounds come from the $n \to 0$ case, for which 
\be
S_0^{\rm c}  =  \frac{4 \pi \ell^2}{27 b^2 n^2} + {\cal O}\(1/n\)\,.
\ee
The $n=0$ constraint thus reads
\be
b_f^2 \leq \frac{b_i^2}{2}\,,
\ee
or equivalently\footnote{For different initial masse this becomes $r_{f} \ge \frac{\sqrt{ (r_{i1}^2 + r_{i2}^2)(\ell^4 + 9 r_{i1}^2 r_{i2}^2)} + \sqrt{12 \ell^2r_{i1}^2 r_{i2}^2+ ( r_{i1}^2 + r_{i2}^2)(\ell^4 + 9 r_{i1}^2 r_{i2}^2)}  }{6  r_{i1}r_{i2} }$ and in the limit $ r_{i1}, r_{i2}\gg \ell$ this reduces to $r_{f} \ge \sqrt{  r_{i1}^2 + r_{i2}^2}$.}
\be
r_{f} \ge \frac{\ell^2 + 3 r_{i}^2 + \sqrt{\ell^4 + 9  r_{i}^4}}{3 \sqrt 2 r_{i}}
\ee
in terms of the event horizons. Substituting into eq.~\eqref{M} this translates into the tightest condition for the final mass
\be \label{eq:Mfn0}
M_f \ge M \(r_+ = \frac{\ell^2 + 3 r_{i}^2 + \sqrt{\ell^4 + 9  r_{i}^4}}{3 \sqrt 2 r_{i}} \) \,.
\ee
 We plot in figure \ref{fig:boundMratio} the ratio
 \be
 \frac{M^{(1)}_{f}}{M^{(0)}_{f}} = \frac{M(r_+ = \sqrt 2 r_{i}) }{M \(r_+ = \frac{\ell^2 + 3 r_{i}^2 + \sqrt{\ell^4 + 9  r_{i}^4}}{3 \sqrt 2 r_{i}} \)}
 \ee
 where the mass denoted $M_f^{(n)}$ is the lower bound on the final mass as predicted from the $n$-th R\'enyi constraint. 
 We see the largest correction to the second law arises for smaller black holes with $r_{i} \sim \ell$ for which ${M^{(1)}_{f}}/{M^{(0)}_{f}}  = \frac 1 6 (64 -19 \sqrt{10}) \approx 0.65$. For $r_{i} \to \infty$ on the other hand, $ {M^{(1)}_{f}}/{M^{(0)}_{f}}  \to 1$. Indeed, in this limit, the inverse temperature $\beta \to 0$ and from eq.~\eqref{S_n}  one sees that the R\'enyi entropies become proportional to the entropy, \ie
 \be
 S_n^{\rm c} (\b) \simeq  \frac{1+ n + n^2}{3 n^2} S(\b) \qquad \mbox{for } \b \to 0\,, \label{eq:Snclargeb}
 \ee
 and hence all of the constraints coincide.

One particularly notable feature of figure \ref{fig:boundM} that both $M^{(0)}_{f}$ and $M^{(1)}_{f}$ exceed the sum of the initial rest masses in the processes. That is, the lower bounds exceed $2 M_i$. To understand how this is possible we need to consider explicitly the total initial energy of the colliding black holes. Let us assume these start at rest, with symmetric initial conditions and collide head-on. The initial energy of the system can then be approximated as a sum of the potential energy within the AdS well and of the Newtonian potential of the two bodies
\begin{align} 
E_i \simeq 2 U_{\rm AdS}+U_{\rm Newton}\,. 
\end{align}
Each black hole starts at $\rho=r/\ell$ away from the origin, with $r$ denoting the position of its centre of mass. Neglecting the backreaction of the masses $M_i$ and Newtonian interaction of the black holes, the potential energy of a point particle of mass $M_i$ sitting at $\rho$ in AdS$_4$ can be written as\footnote{This is obtained from the action of the particle $I =-M\int d\tau \sqrt{-g_{\mu\nu}(x)\partial_\tau x^\mu \partial_\tau x^\nu}$.   Fixing the gauge to $x^0(\tau)=\tau$,  working  in the global AdS coordinates $ds^2  = - \(1+ \frac{r^2}{\ell^2}\) dt^2 + \frac{ dr^2 }{1+ \frac{r^2}{\ell^2} } + r^2  d \Omega^2_{d-1}\, $ and assuming rotational symmetry yields $I=-M\int dt \sqrt{1+\rho^2-\frac{\dot{r}^2}{1+\rho^2}}$.   
Since the Lagrangian is time independent, the conjugate energy is conserved.  We thus evaluate $H=p\dot r - L$ with initial conditions $\dot r=0$ and  get  $H=M\sqrt{1+\rho^2}$.}
\begin{align} 
U_{\rm AdS} =M_i \sqrt{1+\rho^2}\,. 
\end{align}
The Newtonian potential of two such objects in AdS$_4$ instead takes the form
\begin{align} 
U_{\rm Newton}= - \frac{M_i^2}{L}\,, 
\end{align}
where $L$ is the proper distance between the two black holes 
\begin{align} 
L =2 \ell\int_0^\rho \frac{d\rho}{\sqrt{1+\rho^2}} =2 \ell \, \mbox{arcsinh}\rho\, .
\end{align}
All together, we find for the initial energy of the system
\be
E_i \approx 2 \, M_i \sqrt{1+\rho^2} - \frac{M_i^2}{2 \ell  \, \mbox{arcsinh}\rho} \,.
\ee
Of course, this approximation holds as long as $|U_{\rm Newton}| / (2 M_i) \ll 1$. That is,
\be \label{eq:Eregime}
\rho \gg \sinh\(\frac{M_i}{4 \ell}\)\,,
\ee
meaning the initial separation between the colliding black holes is large compared to their size. 
In this regime, $E_i \gg M^{(0)}_{f}\,, M^{(1)}_{f}$ which is then consistent with the results shown in figure \ref{fig:boundM}.


\subsection{Black hole mergers: bounds on gravitational radiation} \label{sec:merg}

To obtain a bound on the energy radiated in the merger, we also consider explicitly the total initial and final energy of the colliding black holes:
\bea
E_i &=& 2 \, M_i \sqrt{1+\rho^2} - \frac{M_i^2}{2 \ell  \, \mbox{arcsinh}\rho} \\
E_f &=& M_f  +E_{\rm rad}\,,
\eea
where $E_{\rm rad}$ denotes the energy radiated gravitationally. 
Energy is conserved in the process and we consider the efficiency with which the total energy $E_i$ is converted into gravitational radiation
\begin{align} 
\epsilon &=\frac{E_{\rm rad}}{E_i} = 1-\frac{M_f}{E_i} \,. 
\end{align}
Notice we are interested in obtaining an efficiency bound at the time of the merger. That is, we are referring to the resulting black hole soon after the merger takes place  and we are ignoring the final fate of the gravitational radiation at late times. The latter is sensitive to the choice of boundary conditions at the AdS asymptotic boundary. For instance with the usual purely reflecting boundary conditions, we expect that all of the gravitational radiation would ultimately be absorbed by the black hole -- see further discussion in section \ref{sec:discuss}.

\begin{figure}[h]
\begin{center}
\includegraphics[width=0.45\textwidth]{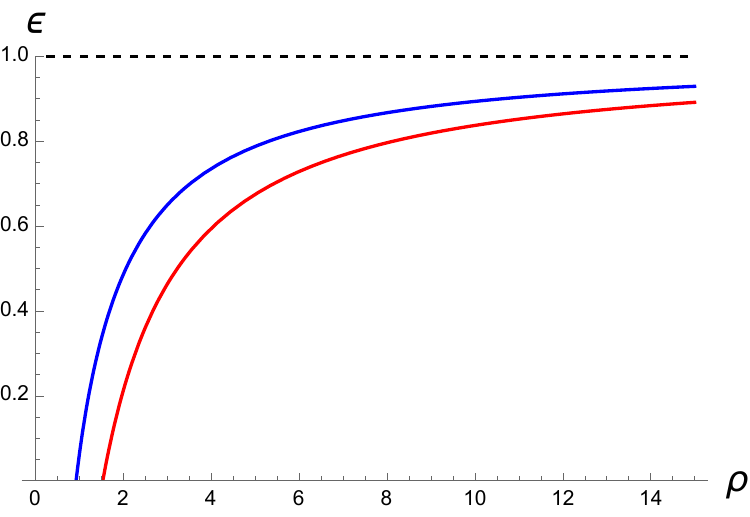} \hfill 
\includegraphics[width=0.45\textwidth]{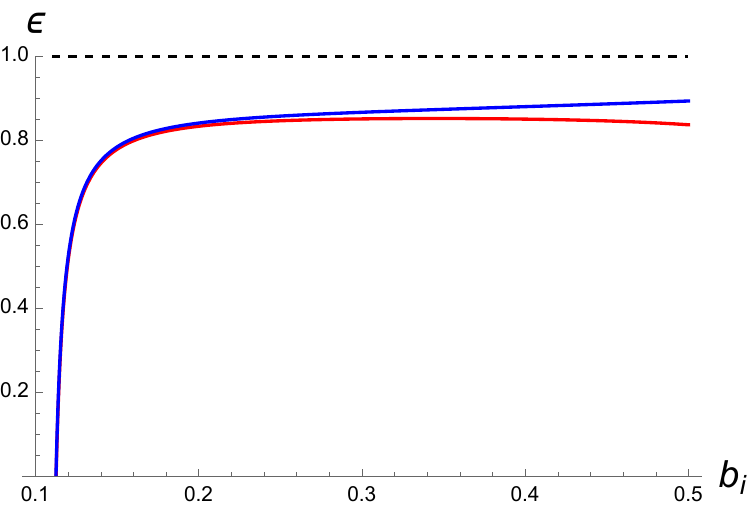} 
\caption{Efficiency upper bounds for symmetric head-on collisions of two AdS$_4$ black holes. As a function of their initial radial position $\rho$, each at the critical  inverse temperature $b_i=1/2$ (left), or as a function of $b_i$ at fixed $\rho = 10$ (right). In blue the $n=1$ bound, in red the $n=0$, in black dashed the maximal theoretical value $\epsilon =1$. }
\label{effrho}
\end{center}
\end{figure}

The area theorem \eqref{eq:Mfn1} then implies the efficiency bound
\begin{align} 
\epsilon^{(1)} & \leq 1 - \frac{M(r_+ = \sqrt 2 r_{i})}{E_i} \,.
\end{align}
However, the most stringent bound comes from the Hartley entropy (\ie the $n\to0$ limit of the R\'enyi entropy) in eq.~\eqref{eq:Mfn0}, which yields
\begin{align} 
\epsilon^{(0)}&\leq 1 - \frac{M \(r_+ = \frac{\ell^2 + 3 r_{i}^2 + \sqrt{\ell^4 + 9  r_{i}^4}}{3 \sqrt 2 r_{i}} \)}{E_i} \,. 
\end{align}
In figure \ref{effrho}, we plot the upper bound of the efficiency for $n=1$ (blue) and $n=0$ (red). In the left panel, we show this bound as a function of the initial positions $\rho$ of the black holes center of mass for a fixed   inverse temperature $b_i = b_\mt{HP} = 1/2$. In the right panel, we have a fixed $\rho$ and a varying function of $b_i$. Notice that the condition \eqref{eq:Eregime} requires in the plots in figure \ref{effrho}  respectively: $\rho \gg 0.25$ (left, $b_i=1/2$) and $b_i \gg 0.23$ (right, $\rho = 10$). 
%


\section{Extremal mergers in AdS$_4$}\label{sec:extremal}

A second remarkably simple scenario is the head-on collision of two extremal black holes. The extremality of the black holes may originate from either charge or spin. These correspond to zero temperature states and, in view of the recent understanding of quantum corrections to extremal solutions, we differentiate between extremal supersymmetric and near-extremal non-supersymmetric black holes when discussing their entropy \cite{Preskill:1991tb,Maldacena:1997ih,Iliesiu:2020qvm,Heydeman:2020hhw,Boruch:2022tno,Iliesiu:2022kny,Iliesiu:2022onk,Sen:2023dps,Ghosh:2019rcj,Rakic:2023vhv,Kapec:2023ruw}. 

As we show below in section \ref{sec:zeroT}, R\'enyi entropies in the supersymmetric case are trivially given by their von Neumann entropies, $S_n(\b \to \infty, J)=S (\b \to \infty, J)$. Thus, if we take these as the initial black holes, no new calculation is required. Moreover, if the final state is a Schwarzschild AdS$_4$ black hole with no charge nor spin, its R\'enyi entropy is again given by eq.~\eqref{S_n}. This allows us to impose new bounds on these processes,  which we study in section~\ref{sec:RenyiExtremal}.
The near-extremal non-supersymmetric case requires instead  a separate treatment due to its vanishingly small degeneracy at low temperatures once one-loop quantum corrections  are taken into account \cite{Preskill:1991tb,Maldacena:1997ih,Iliesiu:2020qvm,Heydeman:2020hhw,Boruch:2022tno,Iliesiu:2022kny,Iliesiu:2022onk,Sen:2023dps,Ghosh:2019rcj,Rakic:2023vhv,Kapec:2023ruw}. We discuss R\'enyi bounds in this instance in section~\ref{sec:nearEx}. 

\subsection{Extremal supersymmetric black holes}

\subsubsection{Zero temperature R\'enyi entropies}\label{sec:zeroT}

Consider for reference the canonical ensemble. 
In the low temperature limit $\beta\rightarrow \infty$, the normalised density matrix acquires the simple form
\begin{align} 
\rho=\frac{e^{-\beta H}}{\tr\, e^{-\beta H}}\underset{\beta\rightarrow \infty}{\longrightarrow} \frac{ \mathbbm 1_N}{N}\,.
\end{align}
Here $N$ refers to the number of non-zero entries of the density matrix (\ie the multiplicity of states of minimal energy), which is nonvanishing for supersymmetric solutions. The von Neumann entropy is then $S=\log N$. The powers of the density matrix give
\begin{align} 
\rho^n=\frac{\mathbbm {1}_N }{N^n} \,,
\end{align}
and thus the R\'enyi entropies equal
\begin{align} 
S_n^{\rm c} (\beta\rightarrow \infty, J) &=\frac{1}{1-n} \log \frac{\tr   \mathbbm {1}_N }{N^n} \nonumber \\
&=\log N \nonumber \\
&=S(\beta\rightarrow \infty, J)\,. 
\end{align}

An alternative route to arrive at the same conclusion is to write eq.~\eqref{eq:Snsecond} in terms of the temperature
\begin{align} 
S_n^{\rm c} (T)&=\frac{n}{(n-1)T}\left( F(T)-F(T/n) \right)
\end{align}
and take the limit $T \to 0$ for fixed $n$. Applying l'Hopital's rule, we easily find
\begin{align} 
S_n^{\rm c}  (T \to 0) &=- \partial_T F |_{T=0}
\end{align}
which is precisely $S(T\to 0)$. Thus, we see that in the extremal supersymmetric limit, all R\'enyi entropies are equal.


\subsubsection{R\'enyi bounds on extremal mergers}\label{sec:RenyiExtremal}

\begin{figure}[h]
\begin{center}
\includegraphics[width=0.5\textwidth]{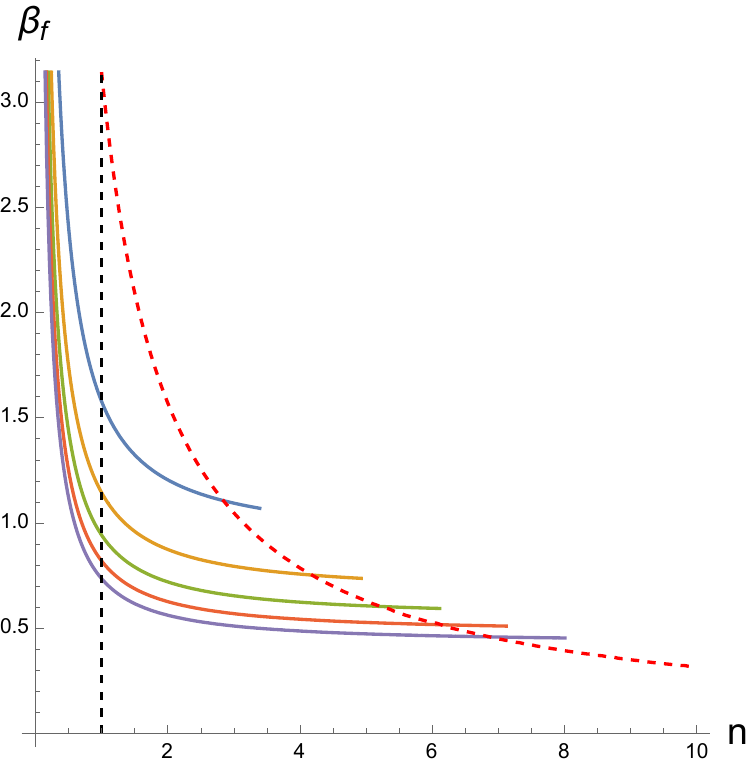} 
\caption{Level curves of $S_n^{\rm c}(\b) = 20, 40, 60, 80, 100$ (where we set $\ell =1$) from the top down for Schwarzschild AdS$_4$. The red dashed curve to the right indicates the Hawking-Page transition of the replicated manifold, beyond which our analysis ceases to be valid. The black dashed vertical line indicates $n=1$.}
\label{Fcontour}
\end{center}
\end{figure}

We consider the simple example of a  black hole merger in which two oppositely rotating or charged extremal BPS AdS$_4$ black holes collide to form a Schwarzschild AdS$_4$ black 
hole.\footnote{Here we mean that individually the two initial black holes would be BPS. Of course, the initial state combining the  two oppositely \eg charged black holes is no longer supersymmetric. However, as in the discussion in section \ref{sec:merM}, if the initial black holes begin at large separation and are prepared independently, we expect that we can evaluate the R\'enyi entropies of the two black holes independently and sum them.} According to our discussion above, the R\'enyi entropy at zero temperature is equal to the thermal entropy and for Schwarzschild AdS$_4$, we have already determined the R\'enyi entropy of the final state, namely eq.~\eqref{S_n}. Therefore, the R\'enyi second laws imply
\begin{align}\label{SfSi}
S_n^{\rm c}  (\b_f) \geq 2 S(\b \to \infty, J)\,.
\end{align}
Again, the simplification of this case lies in that the right hand side of eq.~\eqref{SfSi} is independent of $n$. As a consequence, the task of identifying the strongest bound coming from the family of R\'enyi entropies reduces to an examination of the final state. Given a fixed value of the initial entropy $S(\b \to \infty, J)$, what is the value of $n$ that places the strongest bound on the final mass? Since the mass of the Schwarzschild AdS solution is a monotonic function of its temperature, this is equivalent to determining the strongest bound on the final temperature. The solution is depicted in figure \ref{Fcontour}, where we plot the level curves of $S_n^{\rm c} (\beta)$ from eq.~\eqref{S_n}. 

\begin{figure}[h]
\begin{center}
\includegraphics[width=0.45\textwidth]{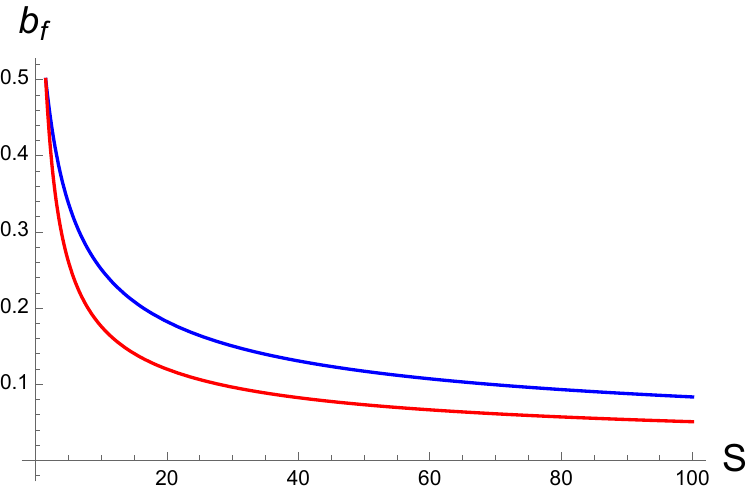} \hfill \includegraphics[width=0.45\textwidth]{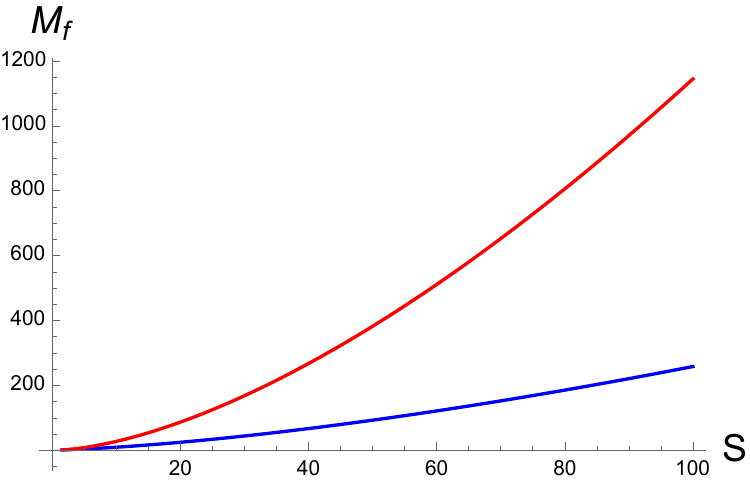} 
\caption{Left: Upper bound on the reduced inverse temperature of the final state. Right: Lower bound on the mass of the final black hole. In both panels $n=1$ (blue) and $n=\frac{1}{2 b_f}$ (red) as a function of the initial entropy $S$. (We have set $\ell=1$ here.) }
\label{fig:extremal}
\end{center}
\end{figure}

In principle, for a given level curve, one should look for the value of $n$ which minimizes the reduced inverse temperature $b_f$ to find the optimal constraint. As we discussed in section \ref{sec:RenyiCanonical}, there is however an extra ingredient: for a given $b$, our analysis ceases to be valid at a critical $n_\mt{HP}  = 1/(2 b)$ because there is a phase transition in the R\'enyi entropies. 
Within the regime of validity of our analysis, the optimal bound thus comes from the limiting value $n_\mt{HP}  =1/(2 b_f)$, for which
\be \label{eq:S12b}
S_{n_\mt{HP} }^{\rm c} (\b)= -  \frac{\pi \ell^2}{27 b^2 (1- 2b)} \left\{ 3 b^2 \(3 +2 \sqrt{1- 3b^2} \) - 2  \(1 + \sqrt{1- 3b^2} \) \right\}\,.
\ee
From Hawking's theorem we have the bound
\be
b_f \leq \frac{2 \ell \sqrt{2 \pi S}}{6 S + \pi \ell^2} \,,
\ee
where for compactness, we denote  $S(\b \to \infty, J) \equiv S$. The most stringent bound similarly comes from substituting eq.~\eqref{eq:S12b} into eq.~\eqref{SfSi}  and solving for $b_f$ the resulting inequality 
\be
- \pi \ell^2 \left\{ 3 b_f^2 \(3 +2 \sqrt{1- 3b_f^2} \) - 2  \(1 + \sqrt{1- 3b_f^2} \) \right\} \ge 54 b_f^2 (1- 2b_f) S\,. 
\ee 
We do not write the explicit expression because of its length, but we plot  the two upper bounds as a function of the initial entropy in figure \ref{fig:extremal}. 

In the small black hole limit, the Hawking bound reaches the critical inverse temperature $b_f = b_\mt{HP} = 1/2$ and  $S = \ell^2 \pi /2$. In this regime the two lower bounds on the mass $M_f$ behave as
\be
M_f - \ell \geq \left\{\begin{array}{ll}
 \frac{2}{\pi \ell} \( S - \frac{\ell^2 \pi}{2}\) + O \( S - \frac{\ell^2 \pi}{2}\)^2 & (n=1) \\
\frac{4}{\pi \ell} \( S - \frac{\ell^2 \pi}{2}\) + O \( S - \frac{\ell^2 \pi}{2}\)^2  & (n=0)
\end{array}\right.  \, . 
\ee
%

\subsection{Near-extremal R\'enyi entropies}\label{sec:nearEx}

Non-supersymmetric near-extremal black holes (in four-dimensional asymtotically flat and AdS spacetimes) have a one-loop canonical partition function, at fixed angular momentum or charge, of the form \cite{Ghosh:2019rcj,Iliesiu:2020qvm,Heydeman:2020hhw,Rakic:2023vhv,Kapec:2023ruw}  
\be \label{Zne}
Z^{\rm c}(\b \gg 1, J) \simeq \b^{-3/2} e^{S_{\rm cl}} \, ,
\ee
with $S_{\rm cl}$ being the classical zero temperature black hole entropy. The corresponding entropy for the near-extremal solution takes the complete form \cite{Rakic:2023vhv,Kapec:2023ruw}
\be \label{eq:NES}
 S^{\rm ne}(\b \gg 1, J) = S_{\rm cl} + a_0 \log S_{\rm cl} + 4 \pi^2 \frac{\b_q}{\b} + \frac{3}{2} \log \frac{\b_q}{\b} + \mathcal O (\b^{-2})\, .
\ee
Here $a_0$ is the numerical coefficient of the logarithmic corrections in the entropy arising from the one-loop determinant of gravitons around the semiclassical black hole saddle \cite{Rakic:2023vhv,Kapec:2023ruw}, but its precise value is not relevant for our discussion.  $\b_q$ denotes an emergent IR scale that in Kerr-AdS$_4$ is directly related the curvature scale of the near-horizon AdS$_2$ factor \cite{Rakic:2023vhv}. 
We have indicated the near-extremal regime as $\beta \gg 1$. More precisely, the expression above is valid in a regime where $\b> \b_q$, but having  $\b< \b_q\, e^{2S_{\rm cl}/3}$ for the one-loop computation to be reliable.

Substituting eq.~\eqref{eq:NES} into eqs.~\eqref{eq:Snsecond} and \eqref{eq:deltaF} to evaluate R\'enyi entropies, we obtain in the near-extremal regime 
 \bea
S_n^{\rm c} (\beta\gg 1, J) &=&  S_{\rm cl} + a_0 \log S_{\rm cl} + \frac{3}{2} \log \frac{\b_q}{\b}  -   \frac{3}{2}\( 1+  \frac{\log n}{1-n}\) + \mathcal O (\b^{-1}) \nonumber \\
 &\simeq &  S^{\rm ne} (\beta\gg 1, J) -  \frac{3}{2}\( 1+   \frac{\log n}{1-n} \) \,. \label{Snearext}
 \eea
 In this case, the zero temperature degeneracy of the R\'enyi entropies of section \ref{sec:zeroT} is lifted by an $n$-dependent but temperature independent term.\footnote{We note that as desired this extra term vanishes in the limit $n\to1$.} Notice however, that contrary to the extremal case, here the entropy $S^{\rm ne}(\b \gg 1, J)$  also has a nontrivial $\b$ dependence.
 
\begin{figure}[th]
\begin{center}
\includegraphics[width=0.5\textwidth]{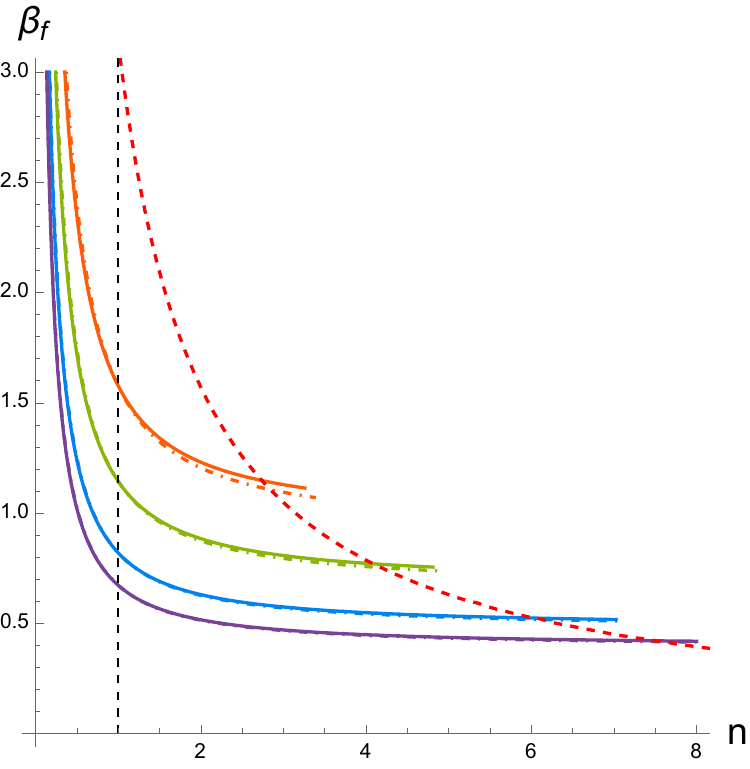} 
\caption{Constraints on the final inverse temperature for the Schwarzschild AdS$_4$ obtained from the merger of two near-extremal Kerr-AdS$_4$ black holes with $  \frac{3}{2} \log \frac{\b}{\b_q} =  \gamma S_{\rm cl}$. These are plotted in solid lines for  $S^{\rm ne}(\b \gg 1, J) \approx (1-\gamma)S_{\rm cl} + a_0 \log S_{\rm cl} = 10,20,40,60$ going from the top down. The dot-dashed lines represent the constraint \eqref{SfSi} which one obtains in the merger of two extremal BPS black holes with  $S(\b \to \infty, J)  =   S^{\rm ne}(\b \gg 1 , J) $ -- see figure \ref{Fcontour}. The red dashed line represents the Hawking-Page transition of the replicated manifold, beyond which our analysis ceases to be valid. In black dashed $n=1$.}
\label{fig:nearExt}
\end{center}
\end{figure}

For a  black hole merger process where two oppositely spinning near-extremal Kerr-AdS$_4$ black holes collide to form a Schwarzschild AdS$_4$ black hole, the R\'enyi second laws require
\be \label{nextR}
S_n^{\rm c}  (\b_f) \geq 2 S_n^{\rm c} (\beta\gg 1 , J)\, .
\ee
These impose an upper bound on the inverse temperature for the final black hole shown in figure~\ref{fig:nearExt}. The qualitative behaviour is similar to the one observed for the extremal case in the previous section (see figure~\ref{Fcontour}), with the most stringent bound coming from the largest allowed value of $n$.  

In producing figure~\ref{fig:nearExt}, we considered near-extremal black holes with $\frac{3}{2} \log \frac{\b}{\b_q} = \gamma S_{\rm cl}$ for some $\gamma <1$, so as to be within the range of validity of eq.~\eqref{eq:NES}. This choice gives   $S^{\rm ne}(\b \gg 1, J) \simeq (1-\gamma)S_{\rm cl} + a_0 \log S_{\rm cl}$ in  eq.~\eqref{Snearext}, and lends itself to a direct comparison with the bounds for the supersymmetric extremal case with matching entropy, \ie with  $S(\b \to \infty, J)$  on the right hand side of eq.~\eqref{SfSi} matching   $S^{\rm ne} (\beta\gg 1, J) $. This isolates the effect of the $\frac{3}{2}\( 1+   \frac{\log n}{1-n} \)$ term in eq.~\eqref{Snearext}. Given that the most stringent bound comes from the largest allowed value of $n$, one sees in  figure~\ref{fig:nearExt}  that its effect is to slightly  relax the constraint as compared to the supersymmetric case. 


\section{Cold to hot mergers in higher dimensional AdS}\label{sec:coldhot}

Another simplification occurs when we restrict to processes where the final state is a Schwarzschild AdS black hole at very high temperature. Here, by high we mean with respect to the AdS scale, \ie $\beta_f\ll \ell$. As we show below, in this regime the R\'enyi entropies acquire a particularly simple form, and become proportional to the entropy $S$ with a prefactor that depends on the R\'enyi index $n$. If additionally we choose initial states that are extremal and supersymmetric, as in section \ref{sec:zeroT}, for which the R\'enyi entropies coincide with $S$, we can easily study the second laws and determine the strongest bound on the evolution. 


\subsection{High temperature R\'enyi entropies}\label{sec:highT}

The R\'enyi entropies acquire a particularly simple form for high temperature AdS black holes. As noted by Hawking \cite{Hawking:1998kw}, the on-shell action for black holes in AdS$_{D}$ far above the Hawking-Page transition is fixed by conformal invariance and scales as
\begin{align}\label{I[b]}
I(\beta)&\sim -\frac{1}{\beta^{D-2}}\,.
\end{align}
The R\'enyi entropies \eqref{eq:Snsecond} then read
\begin{align} 
S_n^{\rm c}(\beta \to 0)&=\frac{1}{1-n} \left( n I(\beta)- I(n\beta) \right) \nonumber \\
&\sim \frac{1}{1-n} \left(-\frac{n}{\beta^{D-2}}+\frac{1}{\left( n\beta \right)^{D-2}} \right) \nonumber \\
&= \left( 1+\frac{1}{n}+\frac{1}{n^2}+\hdots +\frac{1}{n^{D-2}} \right)  \frac{1}{\beta^{D-2}} \,.\label{largeTs}
\end{align}
For $n=1$ this gives $S(\b \to 0) \sim (D-1)/\b^{D-2}$ and thus
\begin{align} 
S_n^{\rm c}(\beta\to 0) &= \frac{1}{D-1}\left( 1+\frac{1}{n}+\frac{1}{n^2}+\hdots +\frac{1}{n^{D-2}} \right) S(\beta\to 0) \nonumber \\
&= \frac{1}{D-1} \frac{1- \frac1{n^{D-1}}}{1-\frac{1}{n}} S(\beta\to 0) \,. \label{largeT}
\end{align}
The above coincides with the high temperature limit \eqref{eq:Snclargeb}, which we worked out explicitly for $D=4$.\footnote{The above expression \reef{largeT} for general $D$ can also be verified using the explicit expressions presented in appendix \ref{app:SAD}.}

An immediate consequence of this factorisation is that, for the collision of very hot Schwarzschild black holes in AdS there is no correction to Hawking's theorem. However, when the colliding states have lower temperature, we do get new constraints. Notice that for BTZ, eq.~\eqref{largeT} is exact for any finite temperature (see appendix \ref{app:AdS3}).

In the limit of large spacetime dimension, eq.~\eqref{largeT} behaves as
\begin{align} 
S_n^{\rm c}(\beta\to 0)\simeq \frac{S(\beta\to 0)}{D} \frac{n}{1-n}  \times
\left\{\begin{array}{ll}
- 1  &\quad  \mbox{for } n>1\,, \\
 \frac1{n^{D-1}} &\quad  \mbox{for } n < 1\,.
\end{array}\right. 
\end{align}
%


\subsection{R\'enyi bounds on cold to hot mergers}

Consider the merger of two opposite extremal BPS black holes.  
Whether extremality comes from rotation or charge is immaterial for this analysis. 
The black holes collide, forming a single black hole and settling at some new inverse temperature $\beta_f$, which we assume to be much smaller than the AdS scale -- this requires of course a large initial angular momentum $J$ or charge $Q$.
The monotonicity constraints of R\'enyi entropies \eqref{SnSn} become
\begin{align} 
S_n^{\rm c}(\beta_f\to 0)&\geq 2 S_n^{\rm c}(\beta \to \infty, J)
\end{align}
and following the arguments above, these take the form
\begin{align}   \label{eq:coldhot}
\frac{1}{D-1} \frac{1- n^{1-D}}{1-\frac{1}{n}} S(\beta_f\to 0) &\geq 2S(\beta \to \infty, J)  \, . 
\end{align}
This expression remains valid as long as $n < n_\mt{HP}$, where as before, $n_\mt{HP}$ denotes the maximum value of $n$ beyond which there is a Hawking-Page transition for the final state on the replicated manifold. 
 In particular for large initial angular momenta (or charge) and thus low final $b_f$, we have $n_\mt{HP}\gg 1$. 
 
 The strongest constraint comes for the value of $n$ that minimises the coefficient in the left hand side of  eq.~\eqref{eq:coldhot}, which is a monotonically decreasing function of $n$. This implies that the most stringent bound, obtained for $n \gg 1$, is   
\begin{align}\label{d-1}
S(\beta_f\to 0)\gtrsim (D-1) 2S(\beta\to \infty, J)\,,
\end{align}
which is $(D-1)$ times the entropy bound predicted by Hawking.

Notice that for $D = 4$,
\be\label{eq:4-1}
S(\beta_f\to 0)\geq  6 S(\beta\to \infty, J)\,,
\ee
and we recover the analysis of section \ref{sec:RenyiExtremal}. In fact for $b_f \to 0$, we have from eq.~\eqref{eq:S12b}
\be
S_{1/(2b)}^{\rm c} (\b_f) \simeq \frac{4 \pi \ell^2}{27 b_f^2} \simeq \frac{S(\b_f)}{3} \ , 
\ee
and using this expression, the condition \eqref{SfSi} coincides with eq.~\eqref{eq:4-1} above. \\

To see what this implies about the final mass $M_f$, recall that the mean energy is computed by
\begin{align} 
M_f =\langle H\rangle=-\partial_{\beta_f} \log Z(\beta_f) \sim \frac{D-2}{\beta^{D-1}}\,,
\end{align}
where we used again eq.~\eqref{I[b]}. Also, as commented above, the entropy of the final state is given by
\begin{align} 
S(\beta_f\to 0)& \sim \frac{D-1}{\beta_f^{D-2}} \,.
\end{align}
Therefore, the mass in terms of the entropy scales as
\begin{align} 
M_f \sim S(\b_f\to 0)^{\frac{D-1}{D-2}}\,,
\end{align}
where we dropped the prefactors. Calling $M_f^{(n)}$ the lower bound on the final mass as predicted from the $n$-th R\'enyi constraint, we find using eq.~\eqref{d-1}
\begin{align} 
\frac{M_f^{(n\to \infty)}}{M_f^{(1)}}= (D-1)^{\frac{D-1}{D-2}}\,.
\end{align}
For example, for $D=3$ (BTZ) the optimal R\'enyi constraint predicts a mass $2^2=4$ times the one predicted by Hawking. For $D=4$, it is $3^{3/2}\simeq 5.2$ times, while, for $D=5$, it is approximately $4^{4/3}\simeq6.3$ times and so on.

Notice that even without considering $n \to \infty$,  if we take eq.~\eqref{eq:coldhot} for $n=2$, we have
\begin{align} 
\frac{M_f^{(2)}}{M_f^{(1)}}= \(\frac{D-1}{2 (1-2^{1-D})}\)^{\frac{D-1}{D-2}}\,.
\end{align}
This evaluates to roughly $1.8,\ 2.2$ and 2.7 respectively for $D=3,\ 4$ and 5.


\section{Additional monotones} \label{sec:repAth}

In this final section, we introduce and discuss another family of monotones that are related to R\'enyi entropies, but descend directly from a notion of entropy in the replica manifold. 

 \subsection{Area theorem in replica manifold}
 
We start from Euclidean semiclassical gravity in AdS, where the black hole solution is associated to the density matrix $\rho$ in a given ensemble.  
As discussed  in section~\ref{ReplicaEnsembles}, considering the $n$ replica  with  appropriate boundary conditions on the boundary   $\partial \mathcal{M}_n$, and solving the vacuum Einstein's equations in the bulk one gets  a smooth geometry $\mathcal{M}_n$  associated to the density matrix
\begin{align} 
 \rho_n=\frac{\rho^n}{\tr  \rho^n }\,. 
\end{align}
At high enough temperatures, above the deconfinement point associated to the replicated manifold, the manifold $\mathcal{M}_n$ possesses a horizon of area $A_n$ with an associated von Neumann entropy given by
\begin{align} 
S \left( \rho_n \right)=\frac{A_n}{4G_N}\,.
\end{align}
Given that $\mathcal{M}_n$ is a well defined and smooth solution to Einstein's equations, and under the same assumptions as Hawking \cite{Hawking:1971tu}, for any transition between a given initial and final state to be possible, these geometries need to satisfy the area theorem constraint
\be \label{area}
S(\rho_n(\lambda))\leq S(\rho_n(\lambda'))\, ,\ \ \ \lambda < \lambda'\,,
\ee
where $\lambda,\lambda'$ is a time (or affine) coordinate in $\mathcal{M}_n$.

Up to this point, we only referred to properties of the manifold $\mathcal{M}_n$ associated to $\rho_n$, but as pointed out in eq.~\eqref{obs}
\begin{align}
S \left( \rho_n \right) &= n^2 \partial_n \left( \frac{n-1}{n} S_n\left( \rho \right) \right) = \tilde S_n\left( \rho \right)\, ,
\end{align}
with $S_n(\rho)$   the $n$-th R\'enyi entropy. This equation links the entropy of $\mathcal{M}_n$, which is associated to an area theorem, to the derivative of the R\'enyi entropies of $\mathcal{M}_1$. 

Then if $t, t'$ is an affine coordinate on $\mathcal{M}_1$, it follows that
\be \label{HR}
\tilde S_n(\rho(t))\leq \tilde S_n(\rho(t'))\, ,\ \ \  t<  t'\,,
\ee
where we have assumed that future-directed is equivalent on $\mathcal{M}_n$ and on $\mathcal{M}_1$, \ie that $t(\lambda)$ is a monotonically increasing function. Eq.~\eqref{HR} provides an additional family of R\'enyi constraints for black holes, which we will refer to as \textit{Hawking-R\'enyi monotones}.

Notice that $\tilde S_n$ may also be rewritten as\footnote{
As discussed in section~\ref{ReplicaEnsembles}, the refined entropy $\tilde S_n$ is ensemble-dependent. Explicitly, we can also here expand eq.~\eqref{eq:SntildeSn} around $n=1$ as we did in eq.~\eqref{eq:dnSn} for the R\'enyi entropy, to write
\bea
\tilde S_n &=& - \frac{\partial_n Z_n|_{n=1}}{Z} + \log Z  + \frac{1-n}{ Z^{2}} \left[ Z \partial^2_n Z_n|_{n=1} - \( \partial_n Z_n|_{n=1}  \)^2 \right] + \hdots \nonumber \\
&=&   S  +\frac{1- n}{Z^2 } \left[ Z \partial^2_n Z_n|_{n=1} - \( \partial_n Z_n|_{n=1}  \)^2 \right] + \hdots \,. 
\eea
The fluctuations are precisely twice those of the R\'enyi entropy in eq.~\eqref{eq:dnSn}. 
}
\begin{align}\label{eq:SntildeSn}
\tilde S_n=S_n+n(n-1)\partial_n S_n \, .
\end{align}
Thus, although in general it is not true that the monotonicity of $\tilde S_n$ implies that of $S_n$, due to the extra derivative term, it is true for two special values: $n=1$ and $n=0$.  In the limit $n \to 1$ in particular, eq.~\eqref{HR} reduces to the standard second law. 


\subsection{Black hole mergers in AdS$_4$}

\begin{figure}[h]
\begin{center}
\includegraphics[width=0.5\textwidth]{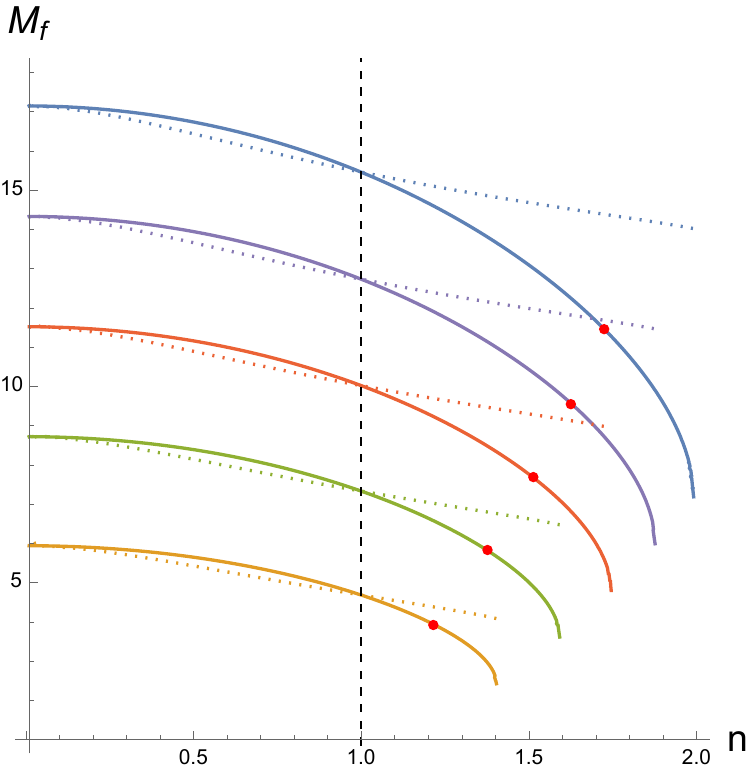} 
\end{center}
\caption{Lower bounds coming from eq.~\eqref{RHboundheadon} on the final mass $M_f$ of a head-on collision in AdS$_4$, as function of the R\'enyi index $n$ (solid lines). Here $M_i = 2,3,4,5,6$ from the bottom up (and we have set $\ell =1$). For comparison we have also reported, in dotted lines, the lower bounds coming from the R\'enyi laws of figure \ref{fig:boundM}. The red dot indicates the value $n_\mt{HP} = 1/(2b_i)$ where a phase transition in the R\'enyi entropies occurs.}\label{fig:RHboundM}
\end{figure}

As a simple application of eq.~\eqref{HR}, we consider again the head-on collision of two identical Schwarzschild AdS$_4$ black holes discussed in section~\ref{sec3+1}. 
Using eqs.~\eqref{S_n} and \eqref{eq:SntildeSn}, the computation of $\tilde S_n$ for a single Schwarzschild black hole in AdS$_4$ is straightforward and leads to
\be\label{Snm}
\tilde S_n^{\rm c}= \frac{\ell^2 \pi }{9 b^2 n^2} \left(1 + \sqrt{1-3 b^2 n^2}\right)^2 \,,
\ee
where $b = \beta / 2 \pi \ell $ is the dimensionless inverse temperature associated with the black hole, related to the mass as in eq.~\eqref{M}:
\be   
M=\frac{\ell}{27 b^3}\left(1+\sqrt{1-3 b^2} \right) \left( 1+ 3 b^2 + \sqrt{1-3 b^2} \right)\,.
\ee
Under the same working assumptions of section \ref{sec3+1}, the Hawking-R\'enyi monotones \eqref{HR} in terms of the initial and final masses imply the following constraints on the evolution:
\be
\tilde S_n^{\rm c} ( M_f) \geq 2 \tilde S_n^{\rm c} (M_i) \,. \label{RHboundheadon}
\ee
In figure \ref{fig:RHboundM}, we plot the lower bound we obtain on the merger's final mass $M_f$ (solid lines). The value predicted by $n=1$, highlighted in the figure, corresponds to the familiar area theorem by Hawking. The mass is measured in AdS units, with $M_i=1$ corresponding to the Hawking-Page transition. It is clear from the figure that the Hawking-R\'enyi inequalities coming from $n<1$ are indeed more constraining than at $n=1$. The optimal bound is achieved at $n=0$, corresponding to the monotonicity of the Hartley entropy $S_0$. Thus, the optimal bound here coincides with the one obtained from the R\'enyi laws studied in section \ref{sec3+1} (reproduced as dotted lines in figure \ref{fig:RHboundM}). However for any given value  of  $n<1$ the Hawking-R\'enyi monotones give a more stringent bound as compared to the corresponding R\'enyi entropies in this process. 
%
 

\section{Discussion} \label{sec:discuss}

\paragraph{Summary:} Any method that provides information about the evolution of a system \textit{without} the need to explicitly solve the equations of motion is highly valuable. In the gravitational context, where Einstein's equations are in general extremely complicated to solve,  Hawking's famous area theorem provides one such method \cite{Hawking:1971tu}. The statement that the horizon area of a black hole is a monotonic non-decreasing function of time imposes strong constraints on the evolution and gives a geometric realization of the second law of thermodynamics.  

As reviewed in section~\ref{sec:RenyiLaws}, there exists in fact a large class of additional monotones that extend the second law and similarly constrain the dynamics. This is well known in the context of quantum thermodynamics, but has only been marginally explored for gravitational systems.  For black holes in AdS spacetime, the focus of this work, a particularly simple --- and computationally accessible ---  set of constraints is provided by the R\'enyi entropies 
\begin{align}
S_n (\r) \equiv \frac{1}{1-n} \log \tr \rho^n\, . 
\end{align}
In contrast to the von Neumann entropy ($n=1$), which takes the same value in every ensemble, these are ensemble-dependent,  as we highlighted in section~\ref{sec:inequiv}. Their non-decreasing behaviour under evolution yields a family of constraints akin to the second law
\be\label{Rineq}
S_n(\rho_i)\leq S_n(\rho_f)\, ,
\ee
\ie a necessary condition to be met for each $n \geq 0$ for a transition between an initial state $\rho_i$ and a final state $\rho_f$. There are of course cases where these are just redundant with the usual second law. This happens for example when the system is described by a microcanonical ensemble or when the entropy is a homogeneous function of the charges, as we discussed in section~\ref{sec:noNew}. However, as we demonstrated in several examples, in general these new entropic monotones do indeed lead to conditions that differ and can be more stringent than the second law.

For all these considerations we needed and implicitly assumed the existence of a well-defined thermodynamic limit. While there may be technical issues with the extent of this assumption for generic gravitational systems (see \eg \cite{Hut:1997wk}), for AdS black holes this limit is guaranteed to exist thanks to holography and the dual CFT description. 

In our analysis, we focused on different scenarios of black hole mergers in AdS. In particular, we considered the head-on collision of two identical black holes of mass $M_i$ resulting in a black hole of mass $M_f$. Under the assumption that the density matrix describing the two colliding black holes approximately factorizes, the R\'enyi constraints \eqref{Rineq}  read
\be \label{RineqM}
S_n (M_f)\geq 2 S_n(M_i)\, . 
\ee 
For each $n$ and fixed $M_i$ this then gives a lower bound on the final mass, in the same way as Hawking's area theorem. The value of $n$ that \textit{maximizes} the bound \eqref{RineqM} provides the physical bound on the final state. Notice that since we are dealing with black holes in AdS, what we refer to as the final state, using the nomenclature of \cite{Hawking:1971tu}, is properly speaking an intermediate (\ie just after merger) state. The subsequent evolution will depend on the precise boundary conditions at the asymptotic boundary. For instance, for purely reflecting boundary conditions, we expect that all gravitational radiation bouncing off from infinity will ultimately be absorbed by the black hole.

The simplest example we analyzed involves  Schwarzschild AdS$_4$ black holes in the canonical ensemble (see appendix~\ref{app:SAD} for the higher dimensional cases). As explained in section~\ref{sec3+1}, R\'enyi entropies can be evaluated explicitly in terms of the temperature, as in eq.~\eqref{S_n}, or equivalently in terms of the mass. However, we noted that eq.~\eqref{S_n} is only valid for a finite range of $n\ (\geq 0$). In fact, for a given $\beta$, at $n_{\mt{HP}}=\pi \ell / \b$, we encounter a phase transition in the R\'enyi entropies due to reaching the Hawking-Page critical point of the replica manifold (see discussion in the final paragraph below eq.~\eqref{S_n}).
Within the range of validity of our analysis, we found that the most stringent constraint does not come from $n=1$. In fact, all conditions with $n<1$ set a larger bound on the mass as compared to the entropy, with $n\to 0 $ giving the strongest constraint (see figure~\ref{fig:boundM}). 
In section~\ref{sec:merg},  we translated the bound on the final mass into a bound on the amount of energy radiated via gravitational waves.  As shown in figure~\ref{effrho}, the R\'enyi  $n=0$ constraint yields an upper bound on the efficiency of energy conversion into radiation that is smaller than the one predicted by the second law. With larger initial black holes (smaller $\beta$) and/or initial separation (larger initial energy), the difference between the two becomes smaller. This is related to having accounted for the initial energy of the system, which we approximated as a sum of the potential energy within the AdS well and of the Newtonian potential of the two colliding bodies. 

In section~\ref{sec:extremal} we instead considered a process where the initial configuration consists of two extremal or near-extremal black holes with opposite spin or charge. 
The extremal configuration corresponds to a zero temperature state and is well understood for supersymmetric black holes. In this case, the R\'enyi entropies coincides with the von Neumann entropy 
\be  \label{susyS}
S^{\rm c}_n(\b \to \infty, J)=S (\b \to \infty, J) = \log N\,,
\ee 
and counts the multiplicity $N$ of the ground state. Taking Schwarzschild AdS$_4$ again as a candidate final state, the R\'enyi constraints  
are nontrivial and show a distinct behaviour with respect to the case of section~\ref{sec3+1}.  The most stringent bound on the final state is now obtained for $n>1$.  More precisely for the limiting value $n=n_{\mt{HP}}=\pi \ell / \b_f$, representing the limit of validity of eq.~\eqref{S_n}.  The dependence of the bound on the entropy characterising the initial state is depicted in figure~\ref{Fcontour}.  

In absence of supersymmetry,  one-loop quantum corrections in the near-extremal regime have been recently understood to render a 
 vanishingly small state degeneracy at low temperatures in the canonical ensemble \cite{Preskill:1991tb,Maldacena:1997ih,Iliesiu:2020qvm,Heydeman:2020hhw,Boruch:2022tno,Iliesiu:2022kny,Iliesiu:2022onk,Sen:2023dps,Ghosh:2019rcj,Rakic:2023vhv,Kapec:2023ruw}. Using the result for Kerr-AdS$_4$ \cite{Rakic:2023vhv,Kapec:2023ruw}, we find that the near-extremal R\'enyi entropies in this case  differ from the corresponding entropy  $ S^{\rm ne} (\beta\gg 1, J)$ by an $n$-dependent term
 \be
S_n^{\rm c} (\beta\gg 1, J) \approx    S^{\rm ne} (\beta\gg 1, J) -  \frac{3}{2}\( 1+   \frac{\log n}{1-n} \) \, .  
 \ee 
As illustrated in figure~\ref{fig:nearExt}, the R\'enyi  constraints are qualitatively similar to those in the supersymmetric extremal case.

The zero temperature initial state with macroscopic degeneracy \eqref{susyS} also lends itself to consider a particular process where full analytic control and an analysis in general dimension is within reach. We considered this in section~\ref{sec:coldhot}, where the final state is represented by a Schwarzschild AdS$_D$ black hole at high temperature. Using the scaling properties of the entropy in the limit $\beta_f\to 0$, one gets 
\be
S_n^{\rm c}(\beta_f\to 0) = \frac{1}{D-1} \frac{1- \frac1{n^{D-1}}}{1-\frac{1}{n}}\, S(\beta_f\to 0) \,,
\ee
and the strongest constraint following from the monotonicity of R\'enyi entropies
\be 
S_n^{\rm c}(\beta_f\to 0) \geq 2 S_n^{\rm c}(\beta \to \infty, J)
\ee
 is obtained from the limiting regime $n\gg1$. 
 
Comparing  the bound on the final mass obtained in this way, $M_f^{(n\to \infty)}$, with the one provided by Hawking's area theorem, $M_f^{(1)}$, yields
\begin{align} 
\frac{M_f^{(n\to \infty)}}{M_f^{(1)}}= (D-1)^{\frac{D-1}{D-2}}\,. 
\end{align}
The R\'enyi laws thus predict a minimal final mass that is a $(D-1)^{\frac{D-1}{D-2}}$ times the one coming from the area theorem.  

These case studies therefore show that: 1) the R\'enyi inequalities \eqref{RineqM} can provide different and more stringent constraints on the allowed configurations when compared to the area theorem, and  2) depending on the specific initial and final configurations, the strongest bound on the final state typically comes either from the smallest or largest value of $n$.

To evaluate  R\'enyi entropies we used the relation to the thermodynamic quantities of the physical state associated with the density matrix $\rho$, as in, \eg section~\ref{sec:RenyiCanonical}. Nonetheless the replica matrix $\rho^n$ entering the definition of $S_n$, or more precisely its normalized version $\rho_n \equiv \frac{\rho^n  }{ \tr \rho^n}$ can also be given a direct geometric interpretation in terms of a replica black hole geometry, as we explained in section~\ref{ReplicaEnsembles}.  This, as the R\'enyi entropies, depends on the ensemble one considers. For a Schwarzschild AdS black hole in the canonical ensemble at temperature $T$, $\r_n$ is represented by the same black hole geometry at temperature $T/n$. 

This geometric representation does not directly connect the monotonicity of $S_n$ to geometric constraints. Still, it prompted us to put forward in section~\ref{sec:repAth} an additional family of constraints that descend from the area theorem in the replica manifold.  The latter implies the monotonicity of entropy associated to the replica state, $\rho_n$.  Noticing then that $S(\rho_n) = \tilde S_n\left( \rho \right)=n^2 \partial_n \left( \frac{n-1}{n} S_n\left( \rho \right) \right)$,  it in turns constrains the original state $\rho$ in terms of the monotonicity of  $\tilde S_n\left( \rho \right)$. 
As shown in figure~\ref{fig:RHboundM}, generally the monotonicity of $\tilde S_n\left( \rho \right)=S(\rho_n)$ and $S_n(\rho)$ produce different conditions at each value of $n$. However, in the example shown there, the strongest constraints both arise at $n=0$ where the two approaches yield the same result.\\

\paragraph{Relation to numerical GR:} 
A primary motivation for studying black hole collisions in general relativity is to describe astrophysical binary mergers, one of the main sources for direct detection of gravitational waves by LIGO and Virgo \cite{LIGOScientific:2016aoc}. This is usually achieved combining analytic and numerical methods \cite{Bantilan:2014sra,Sperhake:2009jz,Zilhao:2012gp,Zilhao:2013nda,Sperhake:2008ga,Bozzola:2022uqu,Shibata:2008rq,Healy:2015mla,Sperhake:2011ik,Sperhake:2010uv,Gold:2012tk,Sperhake:2015siy,Sperhake:2019oaw,Andrade:2019edf,Andrade:2020dgc,East:2012mb,Sperhake:2014nra,Choptuik:2015mma}. While most theoretical modeling resorts to some approximation of the theory, numerical relativity generates solutions to the full non-linear Einstein's equations. In parallel to this direct approach, which is often technically challenging, the second law of black hole thermodynamics poses theoretical constraints that a priori restrict the space of viable processes. However, this bound turns out to be rather loose when compared to full numerical simulations. In four-dimensional flat spacetime, Hawking's area theorem fixes an upper limit of 29\% on the total energy radiated in the merger of two black holes, initially at rest \cite{Hawking:1971tu}. Explicit numerical solutions instead show a value of about 0.1\%, \ie two orders of magnitude smaller than Hawking's bound \cite{Anninos:1993zj}. The area theorem constraint is tighter in ultra-relativistic head-on collisions, where the theoretical upper bound of 29\% \cite{Penrose,Eardley:2002re} is  found to be slightly more than twice the numerical GR result \cite{Sperhake:2008ga}. 

In the present paper, we have studied two families of additional second laws. We explicitly verified they pose more stringent bounds than the area theorem and thus further restrict the space of  physical merging processes. For instance in some of the head-on collisions of two Schwarzschild AdS$_4$ black holes of section~\ref{sec3+1}, the $n=0$ R\'enyi law constraints the maximal efficiency of radiation generation to be about 80\% of the $n=1$ upper bound (see figure~\ref{effrho}). It would be interesting to explore other families of monotones to identify even more stringent bounds that may be informative for actual numerical simulations. 

In this regard, two important remarks are in order: 1) as we stressed, contrary to the standard second law of thermodynamics, the R\'enyi second laws are ensemble dependent, and 2) our analysis here applies only to black holes in AdS space.

As for the first point, in numerical GR simulations, one usually solves Einstein's equations to determine the final equilibrium state by specifying initial data on a Cauchy slice, appropriate boundary conditions and introducing a spacetime discretization. This operational language may appear very different from the statistical ensemble description we can also associate to a given geometry. While the distinction between the two is irrelevant for mean value properties, it becomes important when computing variances and fluctuation-dependent quantities, such as the specific heat, or here the R\'enyi entropies. As emphasized in the main text, the physical predictions of the second laws  \eqref{SnSn} are therefore ensemble dependent. They predict new physical restrictions on the space of solutions in a given statistical ensemble. To make contact with numerical simulations then requires understanding precisely how to relate the choice of boundary conditions throughout the evolution to the statistical description of the system. In the case of charged black holes in AdS for instance the canonical ensemble (at fixed electric charge) differs by the grand canonical ensemble (at fixed electric potential) by an additional boundary term in the total action \cite{Chamblin:1999tk}.

On the other hand, a precise understanding of the appropriate ensemble description for a given simulation is often not available, or beyond the scope of the numerical approach. In those cases, we can instead use the R\'enyi laws to provide information on what is the correct ensemble description for a given process. Imagine  a numerical simulation shows one of the R\'enyi constraint in a given ensemble is violated. Then it means that it is not correct to think of the geometry generated numerically as being described by that ensemble.  

As for the second aspect, we focused on mergers in AdS for a technical reason. To reduce the monotonicity of the R\'enyi divergences  \eqref{monon} to a constraint on R\'enyi entropies \eqref{SnSn}, one needs the maximally mixed state to be a fixed point of the evolution. As we discussed, this is satisfied by black holes in AdS, but not in flat space. Understanding how to evaluate R\'enyi divergences, and possibly other monotones, for asymptotically flat spacetimes, would allow making direct contact with astrophysical mergers (see Outlook paragraph below). 

Still, interestingly, black hole mergers and more general processes of black hole formation in AdS have been extensively studied analytically and numerically in the AdS/CFT literature as a proxy for heavy-ion collisions and quark-gluon plasma formation (see \eg \cite{Bantilan:2014sra,DeWolfe:2013cua,Abajo-Arrastia:2010ajo,Balasubramanian:2010ce,Balasubramanian:2011ur,Balasubramanian:2011at,Allais:2011ys,Balasubramanian:2013rva,Balasubramanian:2013oga,Callebaut:2014tva}). Through the duality, the R\'enyi laws provide new more stringent bounds for those systems as well. More generally, they may also help scrutinize the dynamics of entanglement in equilibration processes at strong coupling.
 
Next to this, in AdS the R\'enyi laws could be used to constrain final states of gravitational processes that are currently beyond reach of analytical approximations and numerical methods, such as the Kerr-AdS superradiant instability \cite{Hawking:1999dp,Cardoso:2004hs,Cardoso:2013pza,Green:2015kur}. For the latter, black resonators \cite{Dias:2015rxy,Chesler:2018txn}, no-end states \cite{Niehoff:2015oga}, multi-oscillatoring black holes \cite{Chesler:2021ehz} and grey galaxies \cite{Kim:2023sig} have been put forward as candidate endpoints of the evolution. Entropic constraints may help discern among them.\\

\paragraph{Connections with other approaches:} An interesting approach towards understanding the second law in the context of AdS black holes was presented in \cite{Engelhardt:2017aux,Engelhardt:2018kcs}. The focus of \cite{Engelhardt:2017aux} was the local notion of an apparent horizon, \ie the outermost compact surface which is marginally outer trapped (on a particular time slice) \cite{Hawking:1973uf}. The discussion is relaxed in \cite{Engelhardt:2018kcs} to consider a certain class of minimal marginal (or ``minmar'') surfaces, which are generalizations of HRT surfaces \cite{Ryu:2006bv,Ryu:2006ef,Hubeny:2007xt,Rangamani:2016dms} and also encompass generic apparent horizons. A key result in \cite{Engelhardt:2018kcs} is that the area (\ie the Bekenstein-Hawking formula) of these minimar surfaces yields the outer entropy,  which results from fixing the exterior geomety and coarse graining over the interior information. Further then, a holographic screen foliated by such minimar surfaces obeys the desired second law. Remarkably, this quantity also has a straightforward interpretation in the boundary theory as the simple entropy, which maximizes the von Neumann entropy constrained to yield the outcomes of certain class of ``simple'' experiments.

Now in our discussions, we are looking at finite transitions of various entropic quantities between an initial and a final state, rather than their continuous evolution as in \cite{Engelhardt:2017aux,Engelhardt:2018kcs}. However, as in the latter, we are considering a coarse grained mixed state for which these entropies increase as the state evolves in time. Of course, our working assumption is that the underlying density matrix describing the individual black holes is approximately given by a particular thermal ensemble. Further, if we consider the von Neumann entropy $S_{n=1}$, we assume that the corresponding spacetime geometry locally resembles that of a stationary black hole and the entropy is given by applying the Bekenstein-Hawking formula to what would be the event horizon of the corresponding isolated black hole. In the context of the black hole merger, these extremal surfaces would correspond to the apparent horzion on the initial and final time slices, matching the minmar surfaces of \cite{Engelhardt:2018kcs}.  Hence our entropy could be seen as the outer entropy above and the corresponding constraint (\eg eq.~\reef{another} with $n=1$) is a particular example of the second law considered there.

Our assumption that initial and final states can be described by  thermal ensembles enables us to extend our analysis beyond the von Neumann entropy. This assumption is crucial to evaluate the corresponding R\'enyi entropies and examine the associated R\'enyi second laws. It would be interesting to explore whether the analysis of \cite{Engelhardt:2017aux,Engelhardt:2018kcs} could be expanded to provide a more precise understanding of the underlying mixed states at general times, which may allow for an evaluation of the corresponding R\'enyi entropies in broader contexts. Additionally, applying their approach to study the evolution of Hawking-R\'enyi monotones, discussed in section \ref{sec:repAth}, may yield fruitful new insights.\\

\paragraph{Outlook:} 
An interesting question we left unexplored regards the \textit{R\'enyi mutual information}. Throughout this work we assumed the density matrices of the two colliding black holes approximately factorize, and discussed in section~\ref{sec3+1} what we expect to be the regime of validity of this approximation.  However, two black holes at a finite distance apart will cause deformations on each other with respect to their states in isolation, which we ignored in our analysis to first approximation. The variation in their respective areas is related to the mutual information, and for sufficiently large black holes these corrections cannot  be neglected anymore. A good starting point from which to evaluate the R\'enyi mutual information in perturbation theory is probably the fully solvable extreme mass ratio merger \cite{Emparan:2016ylg}. \\

In the main text, we mentioned several times that, for a given reduced inverse temperature $b$, the analysis in terms of black hole solutions  is no longer valid for $n>n_{\mt{HP}}=1/(2b)$ since the partition function in the replicated geometry experiences the \textit{Hawking-Page transition}. However, the monotonicity constraints remain valid even in this regime, but with a different interpretation: in this case, we are constraining the evolution from two `copies' of thermal AdS, something reminiscent of the binary merging of two `stars' into a possible final star or black hole. A natural extension of our work would then be a comprehensive analysis of the R\'enyi laws for all gravitational phases. In fact, while we considered here the regime where black holes provide the dominant gravitational saddle, we could similarly evaluate R\'enyi entropies below the Hawking-Page transition from the thermal AdS action.\\

The main focus of our work was to use the information encoded in R\'enyi entropies to pose additional constraints on gravitational dynamics. Having access to the explicit expression of R\'enyi entropies also provides more detailed information about the spectrum of the state described by the density matrix $\rho$. In principle, a knowledge of  $S_n$ for integer $n$ can be used to determine the full \textit{entanglement spectrum} (\ie of the eigenvalues distribution of $\r$) as was explicitly shown in  \cite{Calabrese:2008iby} for one-dimensional discrete systems in the scaling regime. Having obtained general expressions for the R\'enyi entropies for a finite range of the index $n$ here, it is then natural to ask what information about the entanglement spectrum one could extract.
From this point of view, the high temperature result \eqref{largeT} represents a good starting point for an analysis along the lines of that performed in, \eg \cite{Hung:2011nu}. \\

The elegance of the traditional second law lies in its formulation as a geometric principle, Hawking's area increase theorem. The proof of the latter depends on the focusing theorem \cite{Penrose:1968ar,Hawking:1967ju} and the null energy condition \cite{Hawking:1970zqf}, both of which are straightforward local constraints on the geometry and matter. A natural question then arises:\footnote{We thank the referee for raising this interesting question.} do similar local conditions underpin the new R\'enyi second laws, or at least, special cases of these laws for specific values of the R\'enyi index? There may be potential connections to explore here with the quantum focusing conjecture \cite{Bousso:2015mna} and the averaged null energy condition \cite{PhysRevD.17.2521,PhysRevD.44.403,Graham_2007,PhysRevD.54.6233}, though these are not strictly local in the same sense as the former.\\

The thermodynamic instability of flat space black holes prevented us from studying their R\'enyi second laws. In section~\ref{sec:repAth}, however, we proposed an additional family of geometric constraints for AdS black holes, the \textit{Hawking-R\'enyi monotones}. Unlike the R\'enyi second laws, these were not obtained from  monotonicity properties under CPTP maps and do not suffer from issues related to thermodynamic instability (see appendix~\ref{app:flat}). For each $n$, they simply represent the ordinary second law in the replica manifold, following from Hawking's area theorem.\footnote{Notice that black holes in flat space in general fail to satisfy the basic R\'enyi entropies properties in \eqref{ineqs},  but  crucially not the one stating the positivity of $S(\rho_n) = \tilde S_n\left( \rho \right)=n^2 \partial_n \left( \frac{n-1}{n} S_n\left( \rho \right) \right)$ (see comments in app.~\ref{app:flat}).}
This opens up the possibility of extending the analysis also to black holes in asymptotically flat space. We plan to report on this elsewhere in the near future \cite{HRflat}. \\

In the context of Markovian thermal processes, the R\'enyi second laws were shown to be part of a much larger family of entropic monotones known as \textit{h divergences} \cite{Lostaglio:2021voi,Korzekwa:2022wtm}. Further, these works identified within this family the set of most stringent constraints, and solved them in some low-dimensional quantum systems obtaining necessary and sufficient conditions for the existence of a Markovian thermal process between given initial and final energy distributions. This approach essentially trades solving the equations of motion for identifying and solving the `if and only if' set of entropic constraints. In view of our findings, one could speculate whether it may be possibile to extend such an approach to GR.  \\


\section*{Acknowledgements}

We thank Leonardo Banchi, Pasquale Calabrese, Alessandro Cuccoli, William East, Luis Lehner, Jonathan Oppenheim and Erik Tonni for useful discussions and comments at various stages of this work. AB and FG are grateful for the hospitality of Perimeter Institute and of Galileo Galilei Institute for Theoretical Physics (GGI) where part of this work was carried out.  FG  would like to thank the Isaac Newton Institute for Mathematical Sciences, Cambridge, for support and hospitality during the programme Black holes: bridges between number theory and holographic quantum information, where work on this paper was done. This work was supported by EPSRC grant EP/R014604/1. RCM is  supported in part by a Discovery Grant from the Natural Sciences and Engineering Research Council of Canada, and by funding from the BMO Financial Group. 
This research was supported in part by the Simons Foundation through the Simons Foundation Emmy Noether Fellows Program at Perimeter Institute. Research at Perimeter Institute is supported by the Government of Canada through the Department of Innovation, Science and Economic Development and by the Province of Ontario through the Ministry of Research, Innovation and Science. 


\begin{appendix}

\section{R\'enyi entropies for Schwarzschild black holes}

Here we examine the R\'enyi entropies for Schwarzschild black holes in general dimensions $D$. We consider first the case of asymptotically flat  spacetimes and then asymptotically AdS geometries.

\subsection{Schwarzschild black holes}\label{app:flat}

Let us begin by considering the Schwarzschild black hole in asymptotically flat spacetime for $D >3$ dimensions (\eg see \cite{Chamblin:1999tk})
\be
ds^2 = - \( 1- \(\frac{\omega}{r} \)^{D-3} \) dt^2 + \frac{dr^2}{ 1- \(\frac{\omega}{r} \)^{D-3} } + r^2 d\Omega_{D-2}^2\,,
\label{eq:forest}
\ee
with the horizon at $r_+=\omega$. The mass and entropy are given by
\be\label{entrop2}
M = \frac{(D-2)\Omega_{D-2}}{16\pi G_N} \ \omega^{D-3}
\qquad{\rm and}\qquad
S = \frac{A_H}{4 G_N} = \frac{\Omega_{D-2}}{4 G_N}\, \omega^{D-2}\,.
\ee
while the inverse temperature is
\be
\b=\frac{1}{T}=\frac{4\pi  \omega}{D-3}\,. 
\ee
Given eq.~\reef{entrop2}, we see that the entropy is a homogeneous function of the mass with $\nu = \frac{D-2}{D-3}$ following the notation in section \ref{sec:homentropy}, \ie $S(\lambda M)=\lambda^{\frac{D-2}{D-3}} S(M)$. Therefore, from eq.~\eqref{eq:Sn}, the $n$-dependence of the R\'enyi entropy factorizes as
\bea
S_n^{\rm c} (\b) &=& \frac{1}{D-3}  \frac{n \ (1-n^{D-3}) }{1-n} \ S (\b)\,. 
\eea
Notice that this result fails to satisfy the basic properties of R\'enyi entropies \eqref{ineqs} (more precisely,  the second, fourth and fifth inequalities are violated).  As shown in \cite{Hung:2011nu}, this is directly related to the negative specific heat of the asymptotically flat Schwarzschild black hole \reef{eq:forest}, and the resulting instability of its thermal ensemble. Indeed the canonical ensemble is not defined for objects with negative specific heat, as there is no notion of being at equilibrium with a heat bath, and the specific heat is by definition positive in the canonical ensemble since it is the variance of the energy. As a consequence, there also does not exist a thermodynamic limit for these systems, in which the entropy computed in the various ensembles would coincide up to finite shifts. However one can still define entropy in the microcanonical ensemble, and this coincides with the horizon area.

\subsection{Schwarzschild AdS in higher dimensions}\label{app:SAD}

Here, we extend the discussion in section \ref{sec3+1} to Schwarzschild AdS in higher dimensions. The Schwarzschild AdS metric for $D >3$ dimensions is (\eg see \cite{Chamblin:1999tk}) 
\be
ds^2 = - f(r) dt^2 + \frac{dr^2}{f(r)} + r^2 d\O_{D-2}^2 \,,
\ee
where the blackening factor is now given by
\begin{align}\label{fD}
f(r)=1-\left(\frac{\omega}{r}\right)^{D-3}+\left( \frac{r}{\ell} \right)^2\,.
\end{align}
Following eq.~\eqref{entrop2}, the mass and entropy become
\beq\label{entrop3}
S = \frac{A_H}{4 G_N} = \frac{\Omega_{D-2}}{4 G_N} r_+^{D-2}\qquad
{\rm and}\qquad
M = \frac{(D-2)\Omega_{D-2}}{16\pi G_N} \ \omega^{D-3} \,,
\eeq
where $r_+$ corresponds to the position of the event horizon, \ie the real root $f(r=r_+) = 0$. The Hawking temperature can now be written as:
\beq
T= \frac{1}{4\pi } \left( \frac{D-3}{r_+}+(D-1)\,\frac{r_+}{\ell^2} \right) \,. \label{eq:TSchD}
\eeq
For general $D$, the Hawking-Page transition again occurs at $r_+=\ell$  and hence the critical temperature is $T_\mt{HP}=\frac{D-2}{2\pi\ell}$ \cite{Witten:1998zw}. We focus on the larger black holes with $r_+\ge\ell$ or $T\ge T_\mt{HP}$.

As before, to express the free energy as a function of the temperature, we invert eq.~\eqref{eq:TSchD} and take the larger root of large black holes
\begin{align}\label{rp2}
r_+=\frac{\ell}{(D-1) b}\left( 1+\sqrt{1-(D-1)(D-3)\, b^2} \right)\,,
\end{align}
where the dimensionless inverse temperature remains unchanged from eq.~\reef{dimb}, \ie
\begin{align}\label{dimb2}
b \equiv \frac{\b}{2\pi \ell}\,.
\end{align}
Now $b_\mt{HP}=1/(D-2)$ corresponds to the Hawking-Page transition.

In order to evaluate the Renyi entropies, we first combine eqs.~\reef{eq:Snsecond} and \reef{eq:deltaF} to write
\be
S_n^{\rm c} =  \frac{n \b}{1-n} \int_{n \b}^\b \frac{d\tilde \b}{\tilde \b^2} \ S(\tilde \b)\,. 
\ee
However, to evaluate this expression, we follow  \cite{Hung:2011nu} and convert the  integral over $\beta$  to an integral  over the dimensionless horizon radius $x \equiv r_+/\ell$
\be
S_n^{\rm c} =  \frac{n \b}{n-1} \left[  S(x) T(x) \Big|_{x_n}^x - \int_{x_n}^x \frac{d S(\tilde x)}{d \tilde x} T(\tilde x) d \tilde x \right]
\ee
where $x_n$ is the horizon radius corresponding to the inverse temperature $n \beta$. Performing the integral yields 
\be
S_n^{\rm c} =  \frac{n}{n-1} \frac{\Omega_{D-2}\,\beta\,\ell^{D-3}}{16\pi  G_N} \left[ x^{D-3} \(x^2 -1 \)  \right]_{x_n}^x \, .
\ee
In terms of the (dimensionless) inverse temperature $b$, this reads
\be
S_n^{\rm c} =  \frac{n\, b}{n-1} \frac{\Omega_{D-2}  \ell^{D-2}}{8 G_N (D-1)^{D-1} } \[ s( b) - s(nb) \] \label{eq:renD}
\ee
with
\bea
s(b )  &=&  \frac{1}{b^{D-1}}\( - (D-1)^2 b^2   +    \left( 1+\sqrt{1-(D-1)(D-3)\, b^2} \right)^{2}\) \label{eq:sb}\\
&&\qquad\qquad\times\ \ 
\left( 1+\sqrt{1-(D-1)(D-3)\, b^2} \right)^{D-3}\,.\nonumber 
\eea
For a given inverse temperature $b$, this calculation is valid up to  $n = b_\mt{HP}/b = \( (D-2) b\)^{-1}$. 

One can explicitly check that this reproduces the $D=4$ result in eq.~\reef{S_n} and  the high temperature result in eq.~\eqref{largeTs}. Further, taking the limit $n\to1$, one recovers an expression for the entropy found by combining eqs.~\reef{entrop3} and \reef{rp}. Finally, applying the $n\to0$ limit to eqs.~\reef{eq:renD} and \reef{eq:sb} yields 
\be
S_{0}^{\rm c} \simeq  \frac{2^{D-4}\Omega_{D-2}  \ell^{D-2}}{G_N (D-1)^{D-1} (n b )^{D-2} } \,  . \label{eq:sn0}
\ee

\begin{figure}[t]
\begin{center}
\includegraphics[width=0.45\textwidth]{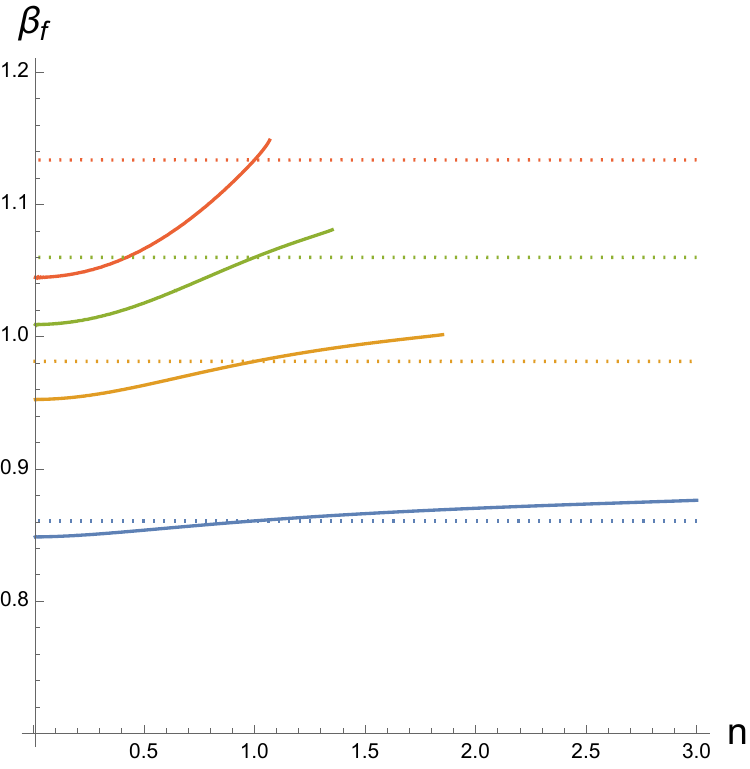} \hfill
\includegraphics[width=0.45\textwidth]{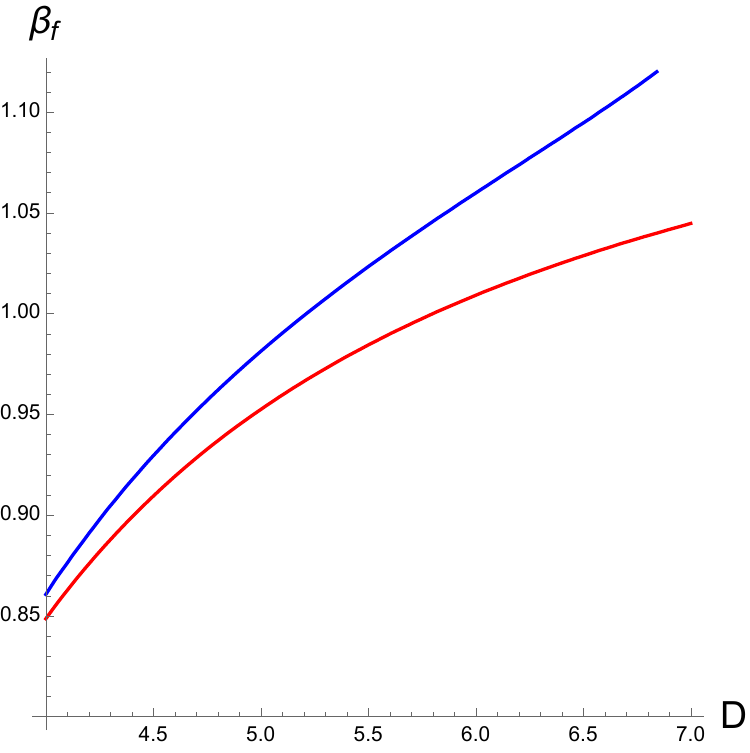} 
\caption{Upper bound on the final inverse temperature for mergers of  Schwarzschild AdS$_D$ black holes with $\beta_i = 1.2$ ($\ell=1$). Left: From the top-down $D = 7,6,5,4$. The dotted lines represent the bound set by the area law ($n=1$). The most stringent bound comes from $n=0$.  Right: The upper bounds on the final black hole inverse temperature onbtained from $n=0$  and $n=1$ as a function of $D$. The $n=0$  (red) constraint is always stronger than the one imposed by the second law $n=1$ (blue).  }
\label{boundsadsD}
\end{center}
\end{figure}

If one then considers a merger of two black holes with equal inverse temperature $\beta_i$ resulting in a single black hole of inverse temperature $\beta_f$, under similar  assumptions to those in section \ref{sec3+1}, the relation 
\be \label{betabound}
S_n^{\rm c} (\beta_f)\geq 2 S_n^{\rm c}  (\beta_i)
\ee
sets an upper bound for the final inverse temperature. Figure~\ref{boundsadsD} shows that again the $n=0$ constraint gives the most stringent  bound.  
To express these results as a bound on the final mass $M_f$, as was done in section \ref{sec3+1}, we must first solve for the horizon radius $r_+$ in terms of the metric parameter $\omega$ which determines the mass in eq.~\reef{entrop3}. This comes from $f(r=r_+) = 0$, which yields
\be
\omega^{D-3}=r_+^{D-3}\(1 + \(\frac{r_+}{\ell}\)^2\)\,,
\ee
but no analytical results are available for general $D$. However, one can still consider special limits (\eg $r_+\to\infty$ for the high temperature limit, as in section~\ref{sec:coldhot}) or evaluate this expression numerically. 
\newline

\begin{figure}[h]
\begin{center}
\includegraphics[width=0.45\textwidth]{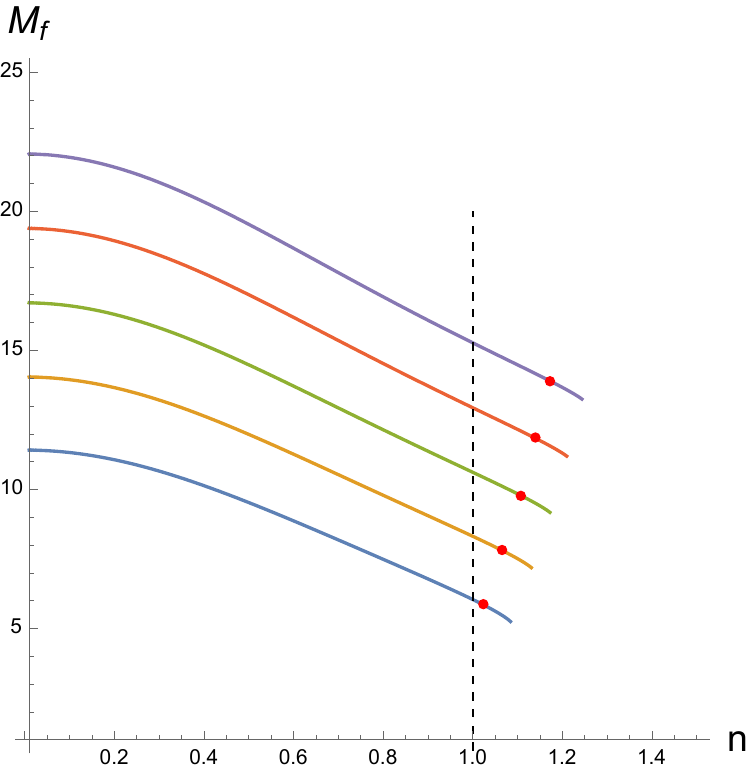}
\end{center}
\caption{Lower bounds on the final mass $M_f$ of a head-on collision in AdS$_5$, as function of the Renyi index $n$. $M_i = 3,4,5,6,7$ from the bottom up and $\ell =1$. Each curve is plotted up to $n = 1/(2 \sqrt 2 b(M_i))$, where the expression \eqref{S_n} becomes complex. In red is marked $n_\mt{HP} =  1/(3b_i)$, which sets the maximal value of $n$ for which eq.~\eqref{S_n} evaluates the R\'enyi entropies.}\label{fig:boundMD5}
\end{figure}

As an example, we show in figure \ref{fig:boundMD5} the results for $D=5$. In this case, the relation  between mass and inverse temperature can be worked out explicitly 
\be 
M = \frac{3 \pi \ell^2}{256 G_N b^4} \left(1 - 4 b^2 + 8 b^4  +  \sqrt{1-8 b^2} \right) .
\ee
and inverted to express eq.~\eqref{betabound} as a condition on the initial and final masses 
\be
S_n^{\rm c}  (M_f)\geq 2 S_n^{\rm c} (M_i)\,. 
\ee
Note the similarity between these results and those shown in figure \ref{fig:boundM} for $D=4$.


\section{BTZ black hole}\label{app:AdS3}


For three-dimensional BTZ black holes, using results from section \ref{sec:homentropy}, we derived eq.~\eqref{eq:RenyigcBTZ} for the grand canonical ensemble
\begin{align} 
S_n^{\rm gc}(\beta,\mu)=\frac{n+1}{2 n}\, S(\beta,\mu)\,. 
\end{align}
Following the same argument, it is also straightforward to show that for a nonrotating BTZ solution, \ie $J=0$, an identical relation holds  in the canonical ensemble 
\begin{align} 
S_n^{\rm c}(\beta)=\frac{n+1}{2 n}\, S(\beta)\,. 
\end{align}

{ Here we make a few important remarks about the Hawking-Page transition in the different ensembles.\footnote{We refer the interested reader to related discussions of the microcanonical ensemble in \cite{Marolf:2018ldl,Marolf:2022jra}.} Consider first the grand-canonical case, where the partition function takes the form $\sum_{M,J}e^{-I_\mt{E}}$ with Euclidean action $I_\mt{E}=\beta M-\mu J-S(M,J)$. One then scans the space of charges $(M,J)$ to find the local minimum $\partial_{M,J}I=0$, and the corresponding solution represents the most likely black hole state with action $I_{\text{BH}}$. Next, one must compare this black hole action to the value of the action $I_{\text{AdS}}$ for pure AdS space (at the same potentials) because this is also an allowed state in the sum. The HP transition occurs when $I_{\text{BH}}=I_{\text{AdS}}$, and in three-dimensions, AdS has charges $M=-1/(8G),J=0$ due to the mass gap. This shows that the HP transition cannot occur in other ensembles such as fixed $(\beta,J\neq 0)$ or $(M>0,\mu)$ as described below, because at least one of the charges is fixed so AdS$_3$ is no longer part of the state sum. 

 }

\subsection{Fixed $(\beta,J)$ ensemble}

A nontrivial example in three dimensions arises when considering a nonstandard ensemble with fixed $\beta$ and angular momentum $J$, for which the partition function reads 
\be
Z (\beta,J) = \tr \ e^{- \beta H }
\ee
with  
\be
I (\beta,J) \equiv -\log Z(\beta,J)  = - \beta M  + S \,.
\ee
Here $\langle H\rangle = M$ and 
\be \label{eq:saddleM}
\beta =   \frac{\partial}{\partial M}S(M,J) \,.
\ee
Equivalently,  working in a saddle point approximation
\bea
Z(\beta,J) &=& \tr \ e^{- \beta H } \\
&=&\sum_{E} e^{-\beta E + S(E,J)}\\
&\approx& e^{-\beta   M+S(M,J)}  
\eea
where  $ M$ now indicates  the saddle, obtained as the solution of  eq.~\eqref{eq:saddleM}.

\begin{figure}[h]
\begin{center}
\includegraphics[width=0.5\textwidth]{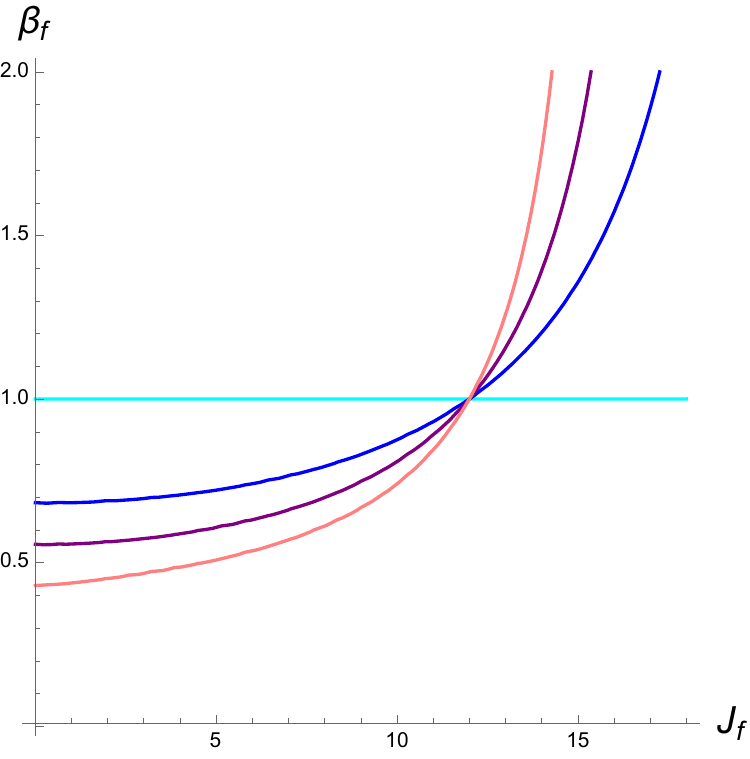} 
\end{center}
\caption{The figure represents various constraints coming from the quantum second laws. Here $G_N = \ell =1$ and the colliding black holes both have $\b_i = 2, J_i = 3$. For each curve, corresponding to a fixed $n$, the states lying above it are forbidden as final states. Only the states lying below all curves are allowed. The colors label $n = 0, 1, 2, 10$ from the top down for  $J_f \to 0$.  } \label{BTZRenyiBetaJ}
\end{figure}

For rotating BTZ black holes with lapse function \eqref{BTZf}, we have \cite{Banados:1992wn}
\be \label{eq:rotatingBTZEH}
r_{\pm} =  2 \ell \sqrt{  G_N  M   \(1 \pm \sqrt{1- \(\frac{J}{M \ell}\)^2}\)}\,,
\ee
in terms of which the thermodynamical variables read
\be
T= \frac{r^2_+ -r^2_-}{2 \pi \ell^2 r_+ }\,, \qquad J = \frac{r_+ r_-}{4 G_N \ell} \,,
\ee
\be
M =\frac{r^2_+ + r^2_-}{8 \pi G_N \ell^2}\,, \qquad  \Omega   =\frac{r_-}{\ell r_+}
\ee
and  
\be \label{eq:rotatingBTZS}
S = \frac{\pi r_+ }{ 2 G_N} =  \ell \pi \sqrt{  \frac{M}{G_N}  \(1 + \sqrt{1- \(\frac{J}{M \ell}\)^2}\)}
\, . 
\ee
 The on-shell action is then given by 
\bea \label{IbJ}
I(\beta,J) &=&  \beta M - S( M, J)  \Big|_{M(\beta,J)}
 \eea
and the R\'enyi entropies are obtained from 
\be
S_n^{\rm \{ \b, J\}} = \frac{1}{1-n}\left[ nI(\b,J)-I(n \b, J) \right] \,.
\ee
We consider the collision of two identical black holes with $(\b_i, J_i  > 0)$ and plot the bounds on the final state $(\b_f, J_f  > 0)$ coming from the R\'enyi constraints in figure \ref{BTZRenyiBetaJ}. 
Only final states that lie below all curves in the $(\b_f, J_f)$ plane are allowed. For all $n$ the curves pass through the point $( \b_f, J_f) = ( \b_i/2,4 J_i)$,\footnote{This can be understood from the scaling properties of the various quantities with the radius. More in detail, using the scaling relation  $4 M(\b, J) = M (\b/2,4 J )$, one verifies that $I( \b_f=\b_i/2,J_f=4 J_i)  =2 I(\b_i,J_i)$, and then $S_n^{\rm \{\beta, J\}} ( \b_f=\b_i/2,J_f=4 J_i) = 2 S_n^{\rm \{\beta, J\}}(\b_i,J_i)$.} making the $n=0$ constraint the most stringent for $J_f \geq 4 J_i$. For $J_f  < 4 J_i$ instead the most stringent constraint comes from $n \to \infty$. { Notice that if the final state has $J_f=0$ the HP transition can occur again, implying that we must restrict to $\beta_f<\frac{2\pi\ell}{n}$. All the curves in this plot satisfy this constraint. This is not relevant since the larger $n$ constraints are suboptimal there, and this process is ruled out due to angular momentum in any event. }  

In particular, the  $n=0$ constraint is $J$-independent as follows from the expression for the  R\'enyi   entropy in the limit $n\to 0$
\be
S_{0}^{\rm \{ \b, J\}} \simeq  \frac{\pi^2 \ell^2}{2 G_N n  \beta } \,,
\ee
which gives the constraint 
\be
\beta_f \leq \frac{\b_i}{2} \, .
\ee
In the opposite limit, $n \to \infty$ one has
\be
S_\infty^{\rm \{ \b, J\}} = - \left[ I(\b,J)- \frac{\b J}{\ell} \right] \,.
\ee
The second term comes from $I(n \b, J)$ and is consistent with the fact that for $n \to \infty$ one is evaluating  $I$ in  eq.~\eqref{IbJ} for an auxiliary black hole in the extremal limit where  $M \ell \to J$ and $S=0$. If one further takes the limit $J \to 0$, one obtains:
\be
S_\infty^{\rm \{ \b, J\to 0\}} = \frac{\pi^2\ell^2 }{2 G_N \beta} \, .
\ee
%

\subsection{Fixed $(M,\mu)$ ensemble}

Another example arises considering an  ensemble with fixed $M$ and angular potential $\mu =\beta \Omega$,  for which the partition function becomes 
\bea
Z(M,\mu) &=& \tr   e^{\mu J} \\
&=&\sum_{J} e^{\mu J+ S(M,J)}\\
&\approx& e^{\mu \tilde J +S( M, \tilde J)} = e^{-I(M ,\mu)} \,,
\eea
and $\tilde J$ solves the saddle point equation 
\be \label{eq:saddleJ}
\mu= - \frac{\partial}{\partial J}S(M,J) \,.
\ee
Using eq.~\eqref{eq:rotatingBTZEH} and \eqref{eq:rotatingBTZS}, 
the solution to eq.~\eqref{eq:saddleJ} fixes  
\be
\tilde J^2 =\ell^2 M^2 \left(1-\left( \frac{2}{ 1  +  \sqrt{ 1+ \frac{16 \mu^2 G_N M}{\pi ^2} } }\right)^2\right)
 \ee
and thus
\be
S(M, \tilde J)  =  \pi \ell  \sqrt{\frac{M}{G_N} \left(1 + \frac{2}{\sqrt{1+\frac{16 \mu^2 G_N M}{\pi ^2}}}\right)}\,.
 \ee
The on-shell action is then given by 
\bea
I(M,\mu) &=&  - \mu \tilde J - S( M, \tilde J) \Big|_{J(M,\mu)}\\
& =&  - \ell  \sqrt{\frac{M}{G_N} \left(1 + \frac{2}{1+\sqrt{1+\frac{16 \mu^2  G_N M}{\pi ^2}}}\right)}\(  \pi + \mu \sqrt{ G_N M \left(1 - \frac{2}{1+\sqrt{1+\frac{16 \mu^2 
G_N M}{\pi ^2}}}\right)} \) \nonumber\\
&&
\eea
and the R\'enyi entropies are obtained from 
\be
S_n^{\rm \{M, \mu\}} = \frac{1}{1-n}\left[ nI(M,\mu)-I(M,n\mu) \right] \,.
\ee
For $n=0$ this is $\mu$-independent and reads
\be
S_0^{\rm \{M, \mu\}} = \pi \ell \sqrt{ \frac{2 M}{G_N}} \, . 
\ee

\begin{figure}[h]
\begin{center}
\includegraphics[width=0.5\textwidth]{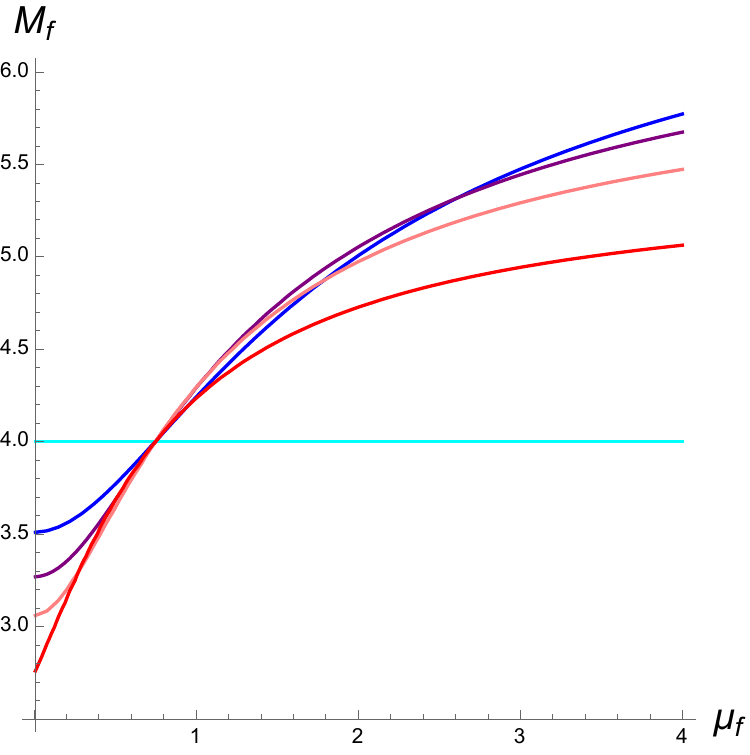} 
\end{center}
\caption{The figure represents various constraints coming from the quantum second laws. Here $\ell =1$ and the colliding black holes both have $M_i = 1, \mu_i = 1.5$. For each curve, corresponding to a fixed $n$, the states lying below it are forbidden as final states. Only the states lying above all curves are allowed. The colors label $n = 0, 0.5, 1, 2, \infty$ from the top down for  $\m_f \to 0$.} \label{BTZRenyiv2}
\end{figure}

We consider the collision of two identical black holes with $(M_i, \m_i  > 0)$ and plot the bounds on the final state $(M_f,\m_f  > 0)$ coming from the R\'enyi constraints in figure \ref{BTZRenyiv2}. 
Only final states that lie above all curves in the $(\m_f, M_f)$ plane are allowed. For all $n$ the curves pass through the point $(\mu_f,M_f) = (\mu_i/2,4 M_i)$,\footnote{Indeed for $(\mu_f,M_f) = (\mu_i/2,4 M_i)$ one has $I(M_f,n\mu_f) =2 I(M_i, n \mu_i)$, and thus $S_n^{\rm \{M, \mu\}}(M_f,\mu_f) = 2 S_n^{\rm \{M, \mu\}}(M_i, \mu_i)$.} making the $n=0$ constraint $M_f \ge 4 M_i$ the most stringent for $\mu_f \leq \mu_i/2$. For $\mu_f  > \mu_i/2$ instead the lower bound is optimized by a different $n$ for each given $\mu_f$, and can be worked out numerically case by case. In the limit $\mu_f \to \infty$, the final state becomes extremal with
 \be
 S_n^{\rm \{M_f, \mu_f \to \infty \}} = \pi \ell \sqrt{\frac{M_f}{G_N}}
 \ee
 becoming $n$-independent. The most stringent R\'enyi bound then comes from taking the $n \to 0$ limit of $S_n^{\rm \{M_i, \mu_i\}}$ and sets $M_f \geq 8 M_i$. 

Notice that also in this case the entropy is a homogeneous function, as in section \ref{sec:homentropy}. The difference resides in the fact that the variables appearing in the entropy are not the conjugate quantities to the thermodynamics variables defining the ensemble.

\end{appendix}


\bibliographystyle{JHEP}
\bibliography{RenyiAdS2}

\end{document}